\documentclass[11pt,a4paper]{article}
\pdfoutput=1
\usepackage{jheppub}
\usepackage{slashed}
\usepackage{braket}

\newcommand{\eq}[1]{eq.~\eqref{eq:#1}}
\newcommand{\eqs}[2]{eqs.~\eqref{eq:#1} and \eqref{eq:#2}}
\renewcommand{\sec}[1]{section~\ref{sec:#1}}

\newcommand{\fig}[1]{figure~\ref{fig:#1}}

\newcommand{\nn}{\nonumber}
\newcommand{\bef}{\begin{figure}[t]\centering}
\newcommand{\eef}{\end{figure}}
\def\bea#1\eea{\begin{align}#1\end{align}}
\def \be  {\begin{equation}}
\def \ee  {\end{equation}}
\def \ba  {\begin{eqnarray}}
\def \ea  {\end{eqnarray}}

\newcommand{\ord}[1]{\mathcal{O}(#1)}

\newcommand{\df}{\mathrm{d}}

\newcommand{\sdt}{\!\cdot\!}

\newcommand{\al}{\alpha}

\newcommand{\de}{\delta}
\newcommand{\eps}{\epsilon}

\newcommand{\si}{\sigma}

\newcommand{\bn}{\bar{n}}

\newcommand{\bnslash}{\bar{n}\!\!\!\slash}

\allowdisplaybreaks[2]

\preprint{\vbox{
\hbox{Nikhef 19-055}}}

\title{The leading jet transverse momentum in inclusive jet production and with a loose jet veto}

\author[a,b]{Darren J.~Scott}
\author[a,b]{Wouter J.~Waalewijn}
\affiliation[a]{Institute for Theoretical Physics Amsterdam and Delta Institute for Theoretical Physics, University of Amsterdam, Science Park 904, 1098 XH Amsterdam, The Netherlands}
\affiliation[b]{Nikhef, Theory Group,
	Science Park 105, 1098 XG, Amsterdam, The Netherlands}
\emailAdd{d.j.scott@uva.nl}
\emailAdd{w.j.waalewijn@uva.nl}

\abstract{
We study the transverse momentum of the leading jet in the limit where the jet radius is small, $R\ll 1$. We introduce the leading-jet function to calculate this cross section for an inclusive jet sample, and the subleading-jet function when a loose veto on additional jets is imposed, i.e.~$p_{T,J} \gtrsim p_T^{\rm veto}$. These jet functions are calculated at next-to-leading order in QCD and the resummation of jet radius logarithms is explored. We present phenomenological results for Higgs + 1 jet production, for both the jet and Higgs transverse momentum distribution. We find that, while the $R \ll 1$ limit of the cross section provides a good description of the full NLO result, even for values as large as $R=0.8$, simply retaining the leading logarithm at this order does not. Indeed, the NLO contribution to the hard function and, to a lesser extent, non-logarithmic corrections to the jet function are sizable and must be included to obtain the correct cross section. In the inclusive cross section we find that the $\al_s^2 \ln^2R$ corrections are several precent, while in exclusive cross sections at large $p_{T,J}$ and small $R$ they can reach 20\%. However, it is not clear how important the resummation of these logarithms is, given the presence of other large corrections at NNLO.
}

\begin{document}
\maketitle

\section{Introduction}
\label{sec:intro}

Jets play a central role in most LHC measurements. The focus is generally on the jet with the largest transverse momentum, referred to as the leading jet. Higher-order QCD corrections to the corresponding cross sections will contain logarithmically enhanced contributions of the form $\al_s^n \ln^m R$, where $m \leq n$, $R$ is the jet radius parameter and $\al_s$ is the strong coupling. Although jet radius logarithms are typically not very large, their study will become increasingly important as experiments use more sophisticated techniques in their analysis, especially when it comes to those involving jet substructure. In fact, experimental studies in this area already use jet radii of $R \sim 0.2$, see e.g.~refs.~\cite{Acharya:2019djg,Aaboud:2019aii}.

We will show how these jet radius logarithms can be resummed to all orders in perturbation theory, through the introduction of a leading-jet function $J_{l,i}(z_l)$. The leading-jet function describes the radiation of a parton $i$ produced in the hard interaction, and how this radiation gets clustered into jets, where $z_l$ is the momentum fraction of the initial parton carried by the leading jet. The properties and number of jets will of course depend on the choice of jet algorithm and jet radius. Clearly however, as one decreases the value of $R$ the probability that emissions from the initiating parton get clustered into separate jets increases, and one might expect a single final state parton to yield two or more jets. We will calculate the jet function at next-to-leading order (NLO) in $\al_s$. Its nonlinear evolution equation resums the logarithms of $R$, and we assess the importance of these corrections in Higgs + 1 jet production. 

The origin of using the small jet radius approximation can be traced back to Sterman and Weinberg~\cite{Sterman:1977wj}, and a first calculation of an inclusive jet cross section in this approximation was carried out in ref.~\cite{Aversa:1989xw}. In exclusive jet cross sections, jet radius logarithms have been resummed for jet rates~\cite{Gerwick:2012fw} and Sterman-Weinberg jets~\cite{Becher:2015hka,Becher:2016mmh}, including a measurement of the jet thrust~\cite{Chien:2015cka}. Jet radius logarithms for exclusive 0-jet cross sections were considered in e.g.~refs.~\cite{Alioli:2013hba, Dasgupta:2014yra, Banfi:2015pju}, and extended to jet mass in exclusive jet production~\cite{Kolodrubetz:2016dzb}. For inclusive jet cross sections, jet radius logarithms have been obtained using the generating functional approach~\cite{Dasgupta:2014yra, Dasgupta:2016bnd}, and through the RG evolution of a jet function~\cite{Kang:2016mcy,Dai:2016hzf} in Soft-Collinear Effective Theory (SCET)~\cite{Bauer:2000ew, Bauer:2000yr, Bauer:2001ct, Bauer:2001yt,Beneke:2002ph}. Our approach will be to follow the latter method to describe the production of the hardest (and second-hardest) jet, and extend the results of ref.~\cite{Dasgupta:2014yra}, by including the hard scattering and also extending it to exclusive one-jet cross sections. 

We will also consider the case where the two jets with the largest transverse momentum are measured, which we refer to as the leading and subleading jet. We describe this in a similar fashion through a subleading-jet function $J_{s,i}(z_l,z_s)$, with $z_l$ and $z_s$ the momentum fractions of the  these jets with respect to the initiating parton. Note that integrating over $z_s$ this simply reduces to the leading-jet function. The subleading-jet function may also be used to describe the production of a single jet with a veto on additional jets, by integrating over $z_s$ up to some cutoff. We will restrict ourselves to the case where the transverse momenta of the jets are of the same order of magnitude, i.e.~$z_l \sim z_s \sim \ord{1}$. For $z_s\ll z_l$ there are additional jet veto logarithms that require resummation, see refs.~\cite{Liu:2012sz,Liu:2013hba}.

The rest of this paper is organized as follows: In \sec{jetf} we define the (sub)leading-jet functions, describe how they enter in factorization theorems, and compute their NLO QCD corrections. In \sec{rge} we derive and solve the nonlinear evolution equation of the (sub)leading-jet function. Finally, in \sec{pheno} we explore the impact of the jet functions in Higgs + 1 jet production and comment on the relative importance of $\ln R$ terms, before concluding in \sec{conc}.

\section{Leading and subleading-jet functions}
\label{sec:jetf}

In \sec{fact}, the (sub)leading-jet functions are introduced, by showing how they enter in factorization theorems. Their field-theoretic definition is given in \sec{def}. The QCD corrections to the (sub)leading-jet functions are calculated at NLO in \sec{nloCalc}.

\subsection{Factorization of the (sub)leading jet cross section}
\label{sec:fact}

We will describe how the (sub)leading-jet function enters in factorization formulae, using Higgs plus inclusive jet production ($HJ$) as a concrete example, which will be studied more closely in \sec{pheno}. We assume that the jet radius is small, i.e.~$R\ll 1$, since our goal is to obtain the (large) jet radius logarithms. This justifies using the collinear approximation in describing the jets produced by the energetic parton(s) exiting the hard scattering process. Starting at leading order (LO) accuracy, the cross section differential in the transverse momentum $p_{T,J}$ of the hardest jet can be written as
\begin{equation} \label{eq:fact_lo}
\frac{\df \sigma^{\rm LO}_{pp\rightarrow HJ}}{\df p_{T,J}} = \sum_{i}\int \! \df p_{T,i}\,\frac{\df \tilde{\sigma}^{(0)}_{pp \rightarrow Hi}}{\df p_{T,i}} \, \int \! \df z_l \, J_{l,i}(z_l,p_{T,i}R, \mu)  \, \delta(p_{T,J} - z_l p_{T,i}) +\ord{R^2}
\,.\end{equation}
The differential cross section $\df {\tilde{\sigma}}^{(0)}_{pp \rightarrow Hi}$ describes the production of a Higgs boson and parton $i$ at leading order in QCD with transverse momentum $p_{T,i}$. As such, it contains the  convolution with initial-state parton distribution functions, along with the usual sum over contributing partonic channels. The sum over $i$ is over possible final-state partons $i=\{q,\bar{q},g\}$. At higher orders, $\df \tilde{\sigma}^{(0)}$  is not simply given by the higher order cross section for Higgs+parton production, as will be discussed shortly, and for this reason we include the tilde.  The jet function $J_{l,i}(z_l,p_{T,i}R, \mu)$ encodes the leading jet produced by this parton with transverse momentum $p_{T,J} = z_l\, p_{T,i}$. In particular, the jet function only depends on the parton initiating the jet and is independent of the process in which it was produced. At LO, the jet function is simply $J_{l,i}(z_l,p_{T,i}R, \mu) = \de(1-z_l)$, so the above equation is trivially correct. However, the factorization in \eq{fact_lo} also holds for the leading logarithms (LL) in $R$. These can be resummed to all orders in $\alpha_s$ by deriving and solving the renormalization group equations (RGEs) for the leading-jet functions, as discussed in \sec{rge}. Thus, \eq{fact_lo} is LO+LL accurate, when including this evolution of the jet function. Note that we assume that there are no other hierarchies in this process, i.e.~$p_{T,J} \sim p_{T,H} \sim m_H$, since these would otherwise introduce additional logarithms requiring resummation. The region $p_T^{\rm veto} \sim p_{T,H} \ll m_H$ has been studied in~\cite{Monni:2019yyr}. 

At NLO, there can be an additional hard parton $j$ (real radiation) in the final state which is not collinear to $i$, so it is necessary to include additional jet functions to describe their radiation\footnote{Factorization formulae where a different number of jet functions appear at different orders in perturbation theory have also been obtained in refs.~\cite{Chang:2013rca,Elder:2017bkd}.}
\begin{align} \label{eq:fact_nlo}
\frac{\df \sigma^{\rm NLO}_{pp\rightarrow HJ}}{\df p_{T,J}} &= \sum_{i,j}\int \! \df p_{T,i}\,\df p_{T,j}\, \frac{\df \tilde{\sigma}_{pp \rightarrow Hij}}{\df p_{T,i}\df p_{T,j}}\, \int \! \df z_{l,i} \, J_{l,i}(z_{l,i},p_{T,i}R, \mu) 
 \\ & \quad \times
\int \! \df z_{l,j} \, J_{l,j}(z_{l,j},p_{T,j}R, \mu)  
 \delta\big(p_{T,J} - \max\{z_{l,i} p_{T,i}, z_{l,j} p_{T,j}\}\big) +\ord{R^2}
\,.\nn \end{align}
Collinear final-state radiation is already encoded by the jet functions and is not part of $\df \tilde{\sigma}_{pp \rightarrow Hij}$, since it would otherwise be double counted.\footnote{Indeed, we extract $\df \tilde{\sigma}_{pp \rightarrow Hij}$ by expanding \eq{fact_nlo} to NLO, and subtracting the contribution involving the NLO jet function. This is discussed in detail in \sec{NLO_sing}.} Furthermore, $\df \tilde{\sigma}_{pp\rightarrow Hij}$ also includes the one-loop virtual corrections to $pp\rightarrow Hi$, for which $p_{T,j}=0$ and the integral over $J_{l,j}$ gives unity from the conservation of probability (see \eq{sumrule}). The cross section $\df \tilde{\sigma}_{pp\rightarrow Hij}$ only depends on the the hard scales $\mu_H \sim p_{T,J} \sim m_H$ and not $R$. As the NLO calculation in \sec{nloCalc} reveals, the jet function on the other hand depends on the scale $\mu_J \sim p_{T,J} R$. The renormalization group evolution between these scales produces the large logarithms of $\mu_J / \mu_H \sim R$, as discussed in \sec{rge}. To extend \eq{fact_nlo} to higher orders in $\al_s$, one needs to include a jet function for each additional well-separated hard parton in the final state. 

We conclude this section with the extension to the case where we are no longer inclusive over the second hardest jet, but instead integrate it up only to some cut-off $p_T^{\rm veto}$. For this, we introduce the subleading-jet function $J_{s,i}(z_l,z_s,p_{T,i}R,\mu)$, 
\begin{align} \label{eq:fact_sub_nlo}
\frac{\df \sigma^{\rm NLO}_{pp\rightarrow HJ}}{\df p_{T,J1}} (p_T^{\rm veto}) & = \int_0^{p_T^{\rm veto}}\! \df p_{T,J2}\,\frac{\df \sigma^{\rm NLO}_{pp\rightarrow HJ}}{\df p_{T,J1} \df p_{T,J2}}
\\ &
= \sum_{i,j}\!\int \!\! \df p_{T,i}\,\df p_{T,j}\, \frac{\df \tilde{\sigma}_{pp \rightarrow Hij}}{\df p_{T,i}\df p_{T,j}}\! \int \! \df z_{l,i} \, \df z_{s,i} \, J_{s,i}(z_{l,i},z_{s,i},p_{T,i} R,\mu) 
\nn \\ & \quad \times
\int \! \df z_{l,j} \, \df z_{s,j} \,J_{s,j}(z_{l,j},z_{s,j},p_{T,j} R,\mu)\,  
 \delta\big(p_{T,J1} \!-\! \max\{z_{l,i} p_{T,i}, z_{l,j} p_{T,j}\}\big) 
 \nn \\ & \quad \times 
 \Theta\big(p_{T}^{\rm veto} \!-\! \max\{\min\{z_{l,i} p_{T,i}, z_{l,j} p_{T,j}\},z_{s,i} p_{T,i}, z_{s,j} p_{T,j}\}\big) 
+\ord{R^2}
.\nn 
\end{align}
In the subleading-jet function, $z_l$ has the same role as in $J_{l,i}$ while $z_s$ denotes the energy fraction carried by the second hardest jet. The theta function describing the second jet is more complicated: If parton $i$ produces the leading jet, the second jet can either be the leading jet produced by parton $j$ or subleading jet from parton $i$. By taking the derivative with $p_T^{\rm veto}$, the spectrum of the second-most energetic jet can also be obtained. This can straightforwardly be extended to also describe the third hardest jet, etc.

\subsection{Definition of the (sub)leading-jet function}
\label{sec:def}

The (sub)leading-jet functions can be defined as collinear matrix elements in SCET, which will be our starting point for their calculation in \sec{nloCalc}. We first introduce  light-cone coordinates 
\begin{align}
  p^\mu = \bn \sdt p\, \frac{n^\mu}{2} + n \sdt p\, \frac{\bn^\mu}{2} + p_\perp^\mu
\,,\end{align}
where $n^\mu = (1,0,0,1) $ is along the parton initiating the jet, $\bn^\mu = (1,0,0,-1)$ and $p_\perp^\mu$ denotes the transverse components. The quark leading-jet function is given by
\begin{align}\label{eq:def_q}
J_{l,q}(z_l, p_T R,\mu) &= 16\pi^3\,\sum_X \frac{1}{2N_c}\, {\rm Tr} \Big[\frac{\bnslash}{2}
\langle 0| \delta(2p_T - \bar n\sdt {\mathcal P}) \delta^2({\mathcal P}_\perp) \chi_n(0)  |X\rangle 
\langle X|\bar \chi_n(0) |0\rangle \Big]
\nn \\ & \quad \times
 \de\bigg(z_l - \frac{\max_{J\in X}  p_{T,J} }{p_T}\bigg) 
\,,\end{align}
where $\chi_n$ is the collinear quark field, $\mathcal P$ the (label) momentum operator, and $N_c = 3$ is the number of colors. Since jet algorithms at hadron colliders are invariant under boosts along the beam axis, it is convenient to work in the frame where the jet has rapidity zero. The $\bar \chi_n(0)$ creates a quark with energy $p_T$, $\chi_n(0)$ annihilates it, and the delta function on the second line picks out the momentum fraction $z_l=p_{T,J}/p_T$ of the leading jet in the state $|X \rangle$. Note that we do not include the top quark and treat the other quarks as massless, so the jet function is independent of flavor. Due to charge conjugation invariance, the quark and anti-quark jet function are the same.
The corresponding definition for gluon jets is given by
\begin{align}\label{eq:def_g}
J_{l,g}(z_l, p_T R, \mu) &= 16\pi^3\,\sum_X \frac{-2p_T}{(d-2)(N_c^2-1)}\,
\langle 0| \delta(2p_T - \bar n\sdt {\mathcal P}) \delta^2({\mathcal P}_\perp) {\mathcal B}_{n\perp}^{\mu,a}(0)  |X\rangle
\nn \\ & \quad \times 
\langle X|{\mathcal B}_{n\perp, \mu}^a(0) |0\rangle
 \de\bigg(z_l- \frac{\max_{J\in X}  p_{T,J} }{p_T}\bigg) 
\,,\end{align}
where ${\mathcal B}_{n\perp}$ is the collinear gluon field and $d=4-2\eps$ in dimensional regularization. For the subleading-jet function an extra delta function must be included for the momentum fraction $z_s$ of the second hardest jet.

We write the perturbative expansion of the leading-jet function as
\begin{equation} \label{eq:J_exp}
J_{l,i}(z_l,p_T R,\mu) = J_{l,i}^{(0)}(z_l,p_T R,\mu) + \left(\frac{\alpha_s}{\pi}\right)J_{l,i}^{(1)}(z_l,p_T R,\mu) + \left(\frac{\alpha_s}{\pi}\right)^2 J_{l,i}^{(2)}(z_l,p_T R,\mu) + \ldots \, ,
\end{equation}
and similarly for the subleading-jet function. The interpretation of these jet functions as a probability implies that they should be normalized to unity
\begin{equation} \label{eq:sumrule}
\int_0^1 \! \df z_l \, \df z_s \, J_{s,i}(z_l,z_s,p_T R,\mu) = \int_0^1 \! \df z_l \, J_{l,i}(z_l,p_T R,\mu) = 1 \, .
\end{equation}
This can be derived from the definitions in \eqs{def_q}{def_g}, and implies, since $J^{(0)}_{l,i}=\delta(1-z_l)$, that the integral of the higher order corrections in the perturbative expansion must vanish,
\begin{equation}
\int_0^1 \! \df z_l \, \df z_s \, J_{s,i}^{(n)}(z_l,z_s,p_T R,\mu) =
\int_0^1 \! \df z_l \, J_{l,i}^{(n)}(z_l,p_T R,\mu) = 0 \qquad \text{for} \ n>0 
\label{eq:sumrule2}
\,.\end{equation}

\subsection{Calculation at next-to-leading order}
\label{sec:nloCalc}

In this section we compute the (sub)leading-jet function up to next-to-leading order (NLO). At LO there are no emissions from the initiating parton. As such there is only one jet, the initial parton. Indeed, evaluating the definitions in \eqs{def_q}{def_g} at this order yields
\begin{equation} \label{eq:lo}
J^{(0)}_{l,i}(z_l,p_T R,\mu) =\delta(1-z_l)\,, \qquad
J_{s,i}^{(0)}(z_l,z_s,p_T R,\mu) = \delta(1-z_l)\delta(z_s) \, .
\end{equation}

We now calculate the one-loop subleading-jet functions in the $\overline{\text{MS}}$ scheme, which can be written as 
\begin{align} \label{eq:jet_nlo}
J_{s,i}^{(1)}(z_l,z_s,p_T R,\mu) &= 
\int\! \mathrm{d}\Phi_2\, \sigma^{c(1)}_{2,i}\,\Big\{\Theta(\theta < R)  \delta(1-z_l)\delta(z_s) + \Theta(\theta > R) \Big[\Theta(x-\tfrac{1}{2})\, \delta(z_l-x)
\nn \\ & \quad
+\Theta(\tfrac{1}{2}-x)\,\delta\big(z_l-(1-x)\big)\Big] \delta\big(z_s-(1-z_l)\big) \Big\}
\,. \end{align}
Here $\theta$ is the angle between the two partons, which have momentum fractions $x$ and $1-x$.
The first term inside the curly brackets corresponds to the case where $\theta$ is less than $R$, so there is a single jet and $z_l = 1$, $z_s = 0$. For the second term, the initiating parton produces two jets, and $z_l$ ($z_s$) is equal to the largest (smallest) momentum fraction. 
Eq.~\eqref{eq:jet_nlo} involves the collinear phase space $\mathrm{d}\Phi_2$ and the one-loop collinear matrix element squared $\sigma^{c(1)}_{2,i}$~\cite{Giele:1991vf}
\begin{align} \label{eq:coll_ME_PS}
\int \mathrm{d}\Phi_2\, \sigma^{c(1)}_{2,q}&=\frac{\alpha_s}{2\pi^{2}} \frac{(e^{\gamma_E}\mu^{2})^{\epsilon}}{\Gamma(1-\epsilon)}\int_0^{2\pi} \mathrm{d\phi} \int_{0}^{1} \mathrm{d}x \; \hat{P}_{qq}(x)\int \frac{\mathrm{d}q_\perp}{q_{\perp}^{1+2\epsilon}}, \nn \\
\int \mathrm{d}\Phi_2\, \sigma^{c(1)}_{2,g}&=\frac{\alpha_s}{2\pi^{2}} \frac{(e^{\gamma_E}\mu^{2})^{\epsilon}}{\Gamma(1-\epsilon)}\int_0^{2\pi} \mathrm{d\phi} \int_{0}^{1} \mathrm{d}x \;\left[n_f \hat{P}_{qg}(x)+\frac{1}{2}\hat{P}_{gg}(x)\right]\int \frac{\mathrm{d}q_\perp}{q_{\perp}^{1+2\epsilon}}
\,.\end{align}
The two partons have transverse momentum $q_\perp$ with respect to the initial parton, which is related to the angle between the two partons by
\begin{align}
  \theta = \frac{q_\perp}{x(1-x)p_T}
\,.\end{align}
The $\hat{P}_{ij}$ in \eq{coll_ME_PS} are given by
\begin{align} \label{eq:P_hat}
\hat{P}_{qq}(x)&= C_F \Big[\frac{1+x^{2}}{1-x}-\epsilon (1-x)\Big], \nn \\
\hat{P}_{qg}(x)&= T_F \big[1-2x(1-x) - 2\epsilon\, x (1-x)\big], \nn \\
\hat{P}_{gg}(x)&= 2C_A \Big[\frac{x}{1-x}+\frac{1-x}{x}+x(1-x)\Big]
.\end{align}
Performing the integrals in \eq{jet_nlo}, we obtain
\begin{align} \label{eq:J_one}
J^{(1)}_{s,q} &= \, \Theta\Bigl(z_l - \frac12\Bigr) \delta(z_s-(1-z_l))\biggl\{\Bigl(\frac{1}{2\epsilon}+\ln\frac{\mu}{p_T R}\Bigr)\bigl[P_{qq}(z_l)+P_{gq}(z_l)\bigr] 
\\ & \quad 
+C_F\biggl[-2\biggl[\frac{\ln(1-z_l)}{1-z_l}\biggr]_+ + \Bigl(\frac{13}{4}-\frac{\pi^2}{3}\Bigr)\delta(1-z_l) - 2\frac{\ln z_l}{1-z_l} 
+ \Bigl(3-\frac{2}{z_l}\Bigr) \ln [z_l(1-z_l)] -\frac{1}{2} \biggr]
\biggr\} 
\,, \nn \\
J^{(1)}_{s,g} &= \Theta\Bigl(z_l - \frac12\Bigr) \delta(z_s-(1-z_l))\bigg\{\Bigl(\frac{1}{2\epsilon}+\ln\frac{\mu}{p_T R}\Bigr)\bigl[P_{gg}(z_l)+2n_f P_{qg}(z_l)\bigr] 
 \nn \\ & \quad 
-2C_A \biggl[\frac{\ln(1-z_l)}{1-z_l}\biggr]_+ - \frac{1}{18}\delta(1-z_l)\bigl[C_A(6\pi^2-67)+23 n_f T_F\bigr] 
 - 2C_A  \frac{\ln z_l}{1-z_l} 
 \nn \\ & \quad
 -\frac{2}{z_l}\bigl[C_A (1-2z_l+z_l^2-z_l^3) + n_f T_F (z_l-2z_l^2+2z_l^3)\bigr]\ln[z_l (1-z_l)] 
- 2n_f T_F z_l (1-z_l)
\bigg\} 
\nn\, .
\end{align}
In \eq{J_one} the regular splitting functions (denoted without a hat) appear, which are
\begin{align*}
P_{qq}(x) &= C_F \bigg(\frac{1+x^2}{[1-x]_+}+\frac{3}{2}\delta(1-x)\bigg) \, , \\
P_{gq}(x) &= C_F  \bigg(\frac{1+(1-x)^2}{x}\bigg) \, , \\
P_{qg}(x) &= T_F \big(x^2 + (1-x)^2\big) \, , \\
P_{gg}(x) &= 2C_A\bigg(\frac{x}{[1-x]_+}+\frac{1-x}{x}+x(1-x)\bigg) + \frac{\beta_0}{2}\delta(1-x) \, , 
\end{align*}
with
\begin{equation*}
\beta_0 = \frac{11C_A-4 T_F n_f}{3} \, .
\end{equation*}

We briefly comment on the form of the subleading-jet functions \eq{J_one}. Since there are at most two partons, the range of $z_l$ is limited to $\tfrac12 \leq z_l \leq 1$ (which extends at order $\al_s^n$ to $\tfrac{1}{n+1} \leq z_l \leq 1$). The $\tfrac{1}{\eps} P_{ij}$ UV divergence leads to an evolution equation involving splitting functions. The structure of this evolution equation is perhaps not immediate and, as we will see in \sec{rge}, is a non-linear DGLAP~\cite{Gribov:1972ri,Altarelli:1977zs,Dokshitzer:1977sg} equation, similar to that for the jet charge~\cite{Waalewijn:2012sv} or fractal observables~\cite{Elder:2017bkd}. We can also read off that the natural scale of these jet functions is $\mu \sim p_T R$, and the evolution to the hard scale $\mu \sim p_T$ will resum the logarithms of $R$.

We end this section by showing numerical results for the one-loop jet functions. In particular, we will  check under what conditions the $\al_s \ln R$ term (the LL term) is a good approximation to the full NLO calculation. At this order in perturbation theory the $z_s$ dependence is completely fixed by $z_l$, so we simply consider the dependence of $J_l$ on $z_l$.
\begin{figure}[t!]
\centering
\includegraphics[width=0.495\textwidth]{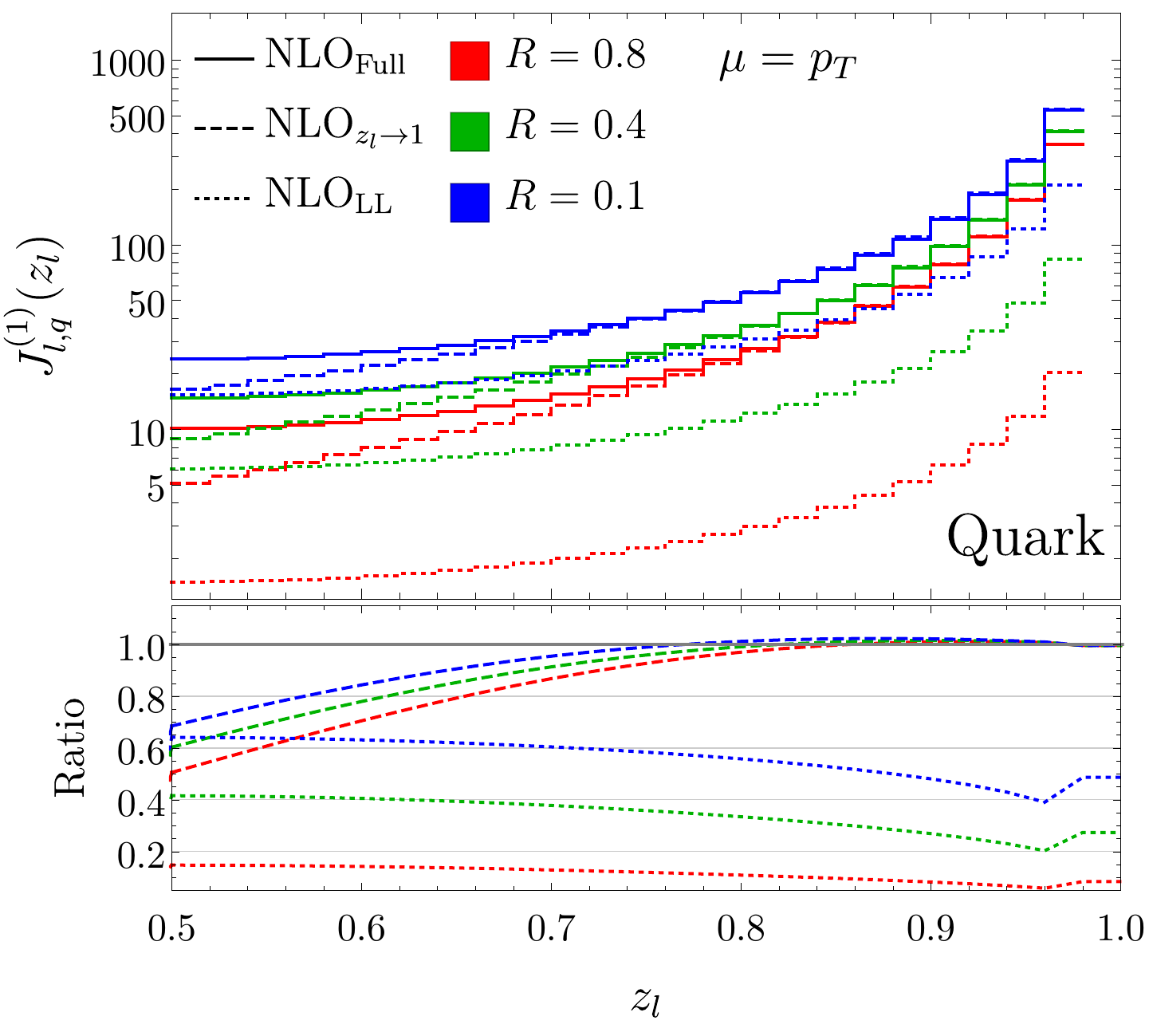}
\includegraphics[width=0.495\textwidth]{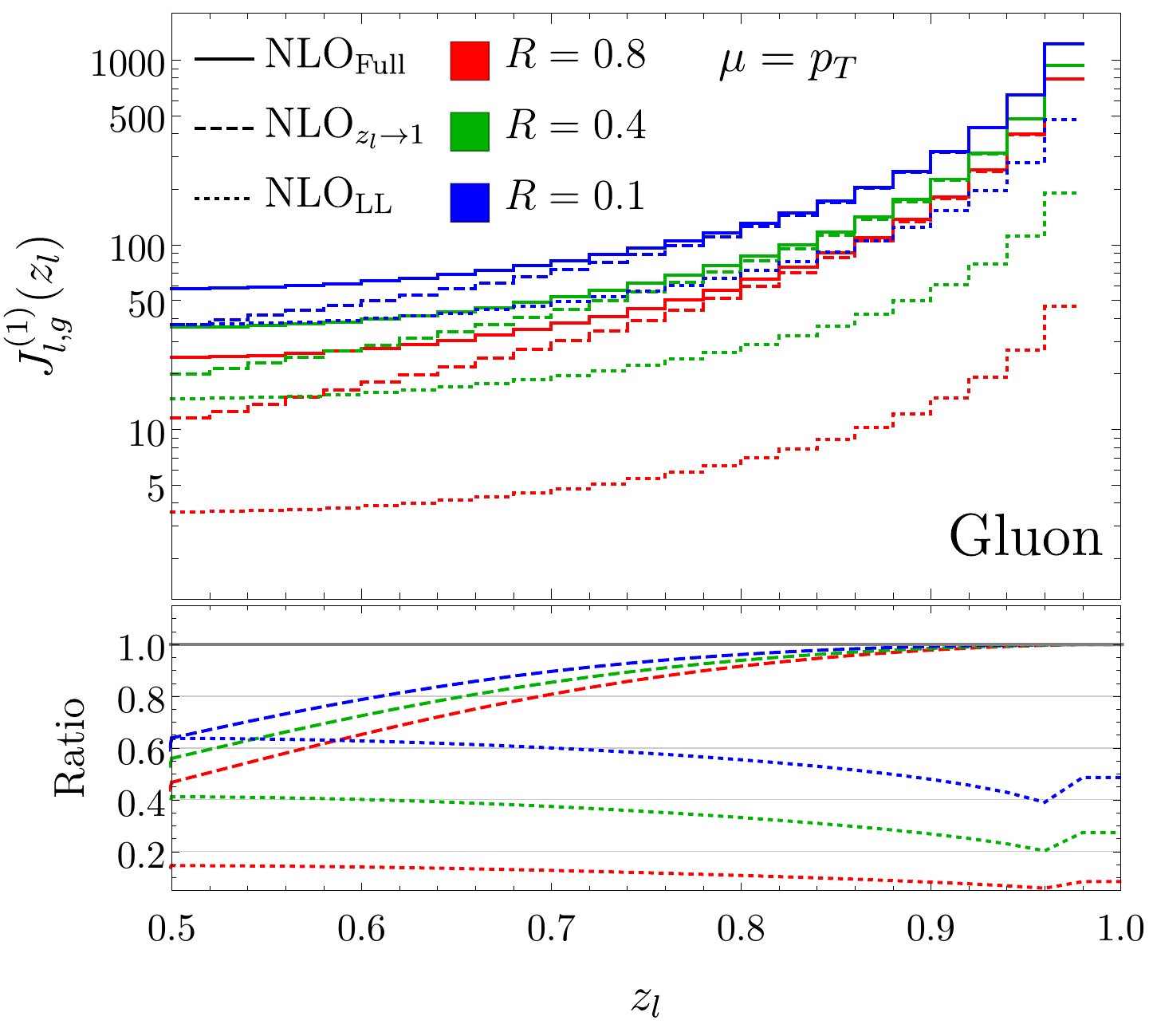}
\caption{The full (solid) leading-jet function at $\ord{\al_s}$, its $\al_s \ln R$ contribution (dotted), and the $z_l \rightarrow 1$ approximation (dashed), for $R=0.8, 0.4, 0.1$ (red, green, blue). Note that the last bin which is negative (and ensures the sum rule in \eq{sumrule2}) does not appear on the log plot.  In the panel below each figure the ratio with the full NLO result is shown.
\label{fig:jFn_NLO}}
\end{figure}
In figure~\ref{fig:jFn_NLO} we show how the quark and gluon leading-jet functions depend on $z_l$ in the left and right panels respectively. To clarify the discussion, we set $\mu=p_T$ so that the jet function involves only $\ln R$ terms and not logarithms involving any other scales. We display each jet function integrated over bins of width $0.02$. The last bin in each plot includes an integration over plus distributions and delta functions and gives a large negative contribution which therefore cannot be seen on the log plot.  Each plot shows the $\al_s \ln R$ contribution (dashed lines) compared to the full NLO result (solid lines) for $R=\{0.8,0.4,0.1\}$. Below each plot we also show their ratio (plotted as a smooth curve between the bins to highlight the trend), indicating that for both the quark and gluon jet functions, the $\al_s \ln R$ term accounts for, at best $60\%$ of the full NLO contribution\footnote{Note that \eq{sumrule} states that the $z_l$ integral  should be  zero. This is not apparent in the plot due to the presence of delta function and plus distributions which give large negative contributions at $z_l=1$.}, and only for very small $R$ values and at moderate values of $z_l \sim 0.5$.

We also examine $z_l \rightarrow 1$ limit of the leading-jet functions, where almost all of the jet momentum is carried by a single parton. This limit is interesting because of the soft singularity of QCD, and the NLO results for the jet functions reduce to
\begin{equation}
\label{eq:J_z1}
J^{(1)}_{l,i} \bigr|_{z_l \rightarrow 1} = \, 2C_i\biggl( \frac{1}{[1-z_l]_+}\ln\frac{\mu}{p_T R} -\biggl[\frac{\ln(1-z_l)}{1-z_l}\biggr]_+ \biggr) + d_i \delta(1-z_l) \, , 
\end{equation}
where $C_q=C_F$, $C_g=C_A$, and
\begin{align}
d_q &= C_F \Bigl(\frac{3}{2}\ln \frac{\mu}{p_T R} + \frac{13}{4}-\frac{\pi^2}{3}\Bigr) \, , \nn \\
 d_g &= \frac{\beta_0}{2}\ln\frac{\mu}{p_T R}+\frac{1}{18}\bigl[C_A(67-6\pi^2)-23n_f T_F\bigr] \, .
\end{align}
These contributions are also shown in figure~\ref{fig:jFn_NLO} (dotted lines). Here we see that indeed, for $z_l>0.7$, these contributions account for $>80\%$ of the value of the jet functions.
Of course, this is only a comparison of the $\ord{\alpha_s}$ corrections, and furthermore these functions appear in cross sections through convolutions with appropriate hard functions. We will investigate the validity of these approximations of the jet functions at the cross section level in \sec{pheno}.

\section{Leading-logarithmic renormalization group equation and solution}
\label{sec:rge}

\begin{figure}[t!]
\centering
\includegraphics[width=0.5\textwidth]{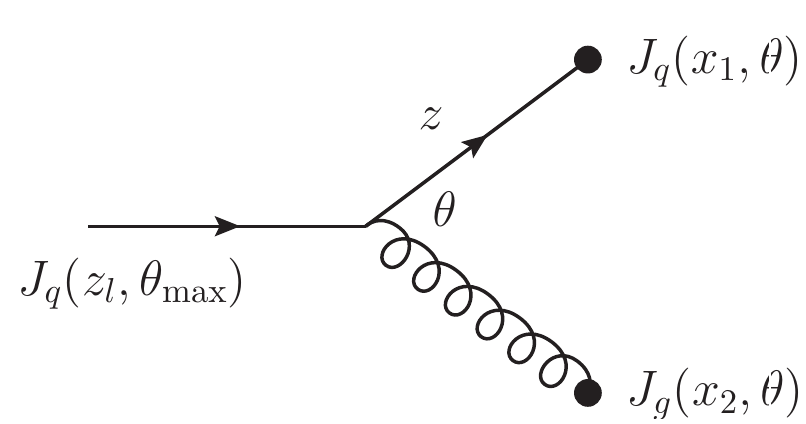}
\caption{\label{fig:lJet_quark}
The leading-jet function of the initial quark can be calculated recursively in terms of the leading-jet functions of the quark and gluon that result from the splitting shown above. Here $z$ and $1-z$ are the momentum fractions of the daughter partons, and $\theta$ is the angle between them.}
\end{figure}

In \sec{rge_derive}, we derive the RGEs for the (sub)leading-jet functions, using a parton shower picture that is accurate to LL order. Our analytical solution to this RGE at order $\al_s^2$ is presented in \sec{rge_solve}, and we also discuss how this can be extended numerically to all-orders in $\al_s$.

\subsection{Deriving the renormalization group equation}
\label{sec:rge_derive}

At LL accuracy, we can think of the radiation produced by a parton as a tree of subsequent $1 \to 2$ splittings that are strongly ordered in angle, i.e.~the angle of subsequent splittings are parametrically smaller. The RGE can now be derived by considering a single $1 \to 2$ splitting, pictured in \fig{lJet_quark}, and noting that the momentum fraction of the leading jet can be calculated recursively (see also ref.~\cite{Elder:2017bkd}). Specifically, the leading-jet function for the initial parton can be obtained from the leading-jet functions of the two daughter partons, by accounting for the distribution in the momentum fractions $z$ and $1-z$ and angle $\theta>R$ of the daughters. Denoting the momentum fraction of the leading jet produced by the daughter partons as $x_1$ and $x_2$, the momentum fraction of the leading jet produced by our initial parton is simply given by
\begin{equation}
z_l = \max\bigl[z x_1, (1-z)x_2 \bigr] \, .
\end{equation}
The additional factors of $z$ and $1-z$ enter because the momentum fractions $x_i$ are of the leading jet produced by a daughter with respect to the momentum of a daughter.

This allows us to immediately write down a recursive expression for the leading-jet function at LL accuracy. For an initial quark this reads 
\begin{align} \label{eq:J_l_deriv}
J_{l,q}(z_l,p_T R,p_T\theta^{\text{max}})&=\delta(1-z_l)+\frac{\alpha_s}{\pi}\int_R^{\theta^{\text{max}}}\frac{\df \theta}{\theta} \int_0^1 \df z\,  P_{qq}(z)\, \int_0^1 \! \df x_1 \, \df x_2 \\
& \quad \times J_{l,q}(x_1,p_TR, p_T \theta)\,J_{l,g}(x_2,p_TR, p_T \theta)\, \delta\big(z_l - \max[z x_1,(1-z)x_2]\big)\,. \nn
\end{align}
The expression for an initiating gluon is very similar, with the appropriate replacements for the splitting and jet functions, except that both $g \rightarrow gg$ and $g \rightarrow q\bar{q}$ splittings contribute. To account for the angular ordering, the jet functions carry the additional argument $\theta$, which gives the \emph{largest} angle at which the next splitting can occur. For the initial splitting $\theta^{\rm max}$ is assumed to be an order one number. For subsequent splittings the upper limit on the $\theta$ integral is set by the angular ordering condition, while the lower limit is always set by the jet radius parameter $R$. To derive the RGE, it is useful to make a change of variables from $\theta$ to the transverse momentum $q_\perp$ of the daughter parton with respect to the initial parton 
\begin{equation}
\label{eq:cov_theta}
\int_R^{\theta^\text{max}}\frac{\df \theta}{\theta} = \int^{p_T}_{p_T R}\frac{\df q_\perp}{q_\perp}=\int^{\mu}_{\mu_0}\frac{\df q_\perp}{q_\perp} \, .
\end{equation}
Note that this transverse momentum is the scale at which $\al_s$ in \eq{J_l_deriv} is evaluated (treating $x_i$ as order one numbers). Taking the derivative with respect to $\mu$, we obtain the RGE. From the lower limit of the integral in \eq{cov_theta} we also read off the natural scale of the jet functions to be $p_T R$. 

For a parton of type $i$, we find for the leading-jet function 
\begin{align}
\label{eq:RG_lJetFn}
\mu \frac{\df}{\df \mu}J_{l,i}(z_l,\mu) &= \frac{\alpha_s(\mu)}{\pi} \int_0^1 \! \df z \, \df x_1 \, \df x_2 \, K_{l,i}(x_1,x_2,z;\mu)\, \delta\big(z_l-\max{[z x_1, (1-z)x_2]}\big) \, ,
\end{align}
where
\begin{align}
K_{l,q}(x_1,x_2,z;\mu) & = P_{qq}(z)J_{l,q}(x_1,\mu) J_{l,g}(x_2,\mu) \, , \nn \\
K_{l,g}(x_1,x_2,z;\mu) & = \tfrac12 P_{gg}(z)J_{l,g}(x_1,\mu) J_{l,g}(x_2,\mu) + n_f P_{qg}(z)J_{l,q}(x_1,\mu) J_{l,q}(x_2,\mu) \, .
\end{align}
Here and in the following we suppress the dependence on the scale $p_T R$ in the arguments of each of the jet functions, to keep the notation compact.

We can similarly derive the LL accurate RGE for the subleading-jet function. We find 
\begin{align} \label{eq:RG_slJetFn}
\mu \frac{\df}{\df \mu}J_{s,i}(z_l,z_s,\mu) &= \frac{\alpha_s(\mu)}{\pi}\int_0^1 \! \df z \, \df x_{1l} \, \df x_{2l} \, \df x_{1s} \, \df x_{2s} \, K_{s,i}(x_{1l},x_{2l},x_{1s},x_{2s},z;\mu) \\
& \quad \times \bigg\{\Theta\big(z x_{1l}-(1-z)x_{2l}\big) \delta \big(z_l - z x_{1l}\big) \delta\big(z_s-\max{\left[z x_{1s},(1-z)x_{2l}\right]}\big) \nn \\
&\quad + \Theta\big((1\!-\!z)x_{2l}\!-\!z x_{1l}\big) \delta\big(z_l\!-\!(1\!-\!z)x_{2l}\big) \delta\big(z_s\!-\!\max{\left[z x_{1l},(1\!-\!z)x_{2s}\right]}\big) \biggr\}\,, \nn
\end{align}
where
\begin{align}
K_{s,q}(x_{1l},x_{2l},x_{1s},x_{2s},z;\mu) &= P_{qq}(z) J_{s,q}(x_{1l},x_{1s},\mu)J_{s,g}(x_{2l},x_{2s},\mu) \, , \nn \\
K_{s,g}(x_{1l},x_{2l},x_{1s},x_{2s},z;\mu) &=\tfrac12 P_{gg}(z) J_{s,g}(x_{1l},x_{1s},\mu)J_{s,g}(x_{2l},x_{2s},\mu) 
\nn \\ & \quad
+ n_f P_{qg}(z) J_{s,q}(x_{1l},x_{1s},\mu)J_{s,q}(x_{2l},x_{2s},\mu) \, .
\end{align}
Note that the constrains $x_{jl}\geq x_{js}$ and $x_{jl}+x_{js}\leq 1$ for $j=1,2$ are automatically satisfied by the jet functions and follow from their definition.

\subsection{Solving the renormalization group equation}
\label{sec:rge_solve}

We now use \eq{RG_lJetFn} to generate higher order terms in $\al_s$ for the LL solution of the leading-jet functions from lower order ones. It is straightforward to check that inserting the LO solutions $J_{l,i}^{(0)}(z_l,\mu)=\delta(1-z_l)$ into \eq{RG_lJetFn} yields the LL part of $J_{l,i}^{(1)}$ in \eq{J_one},
\begin{align}
J_{l,q}^{(1)\rm{LL}}(z_l,\mu)&=\ln\Bigl(\frac{\mu}{p_T R}\Bigr)\,\Theta\Bigl(z_l - \frac12\Bigr)\bigl[P_{qq}(z_l)+P_{gq}(z_l)\bigr] \, ,  \nn \\
J_{l,g}^{(1)\rm{LL}}(z_l,\mu)&=\ln\Bigl(\frac{\mu}{p_T R}\Bigr)\,\Theta\Bigl(z_l - \frac12\Bigr)\bigl[P_{gg}(z_l)+2n_f P_{qg}(z_l)\bigr] \, .
\end{align}

Inserting $J_{l,i}^{(0)}+ \frac{\alpha_s}{\pi}  J_{l,i}^{(1)\rm{LL}}$  into \eq{RG_lJetFn} and expanding everything to order $\alpha_s^2$ we obtain the LL terms at order $\al_s^2$. For the quark leading-jet function these are 
\begin{align}
\label{eq:lJet_q_NNLO}
J_{l,q}^{(2)\rm{LL}}(z_l,\mu) &= \ln^2\Bigl(\frac{\mu}{p_T R}\Bigr)\biggl\{\Theta\Bigl(z_l-\frac12\Bigr) \frac{\beta_0}{4}\bigl[P_{qq}(z_l)+P_{gq}(z_l)\bigr] \nn \\
& \quad + \Theta\Bigl(z_l-\frac12\Bigr)\bigl(C_F^2 A_{q,1} + C_F C_A A_{q,2} + C_F n_f T_F A_{q,3}\bigr) \nn \\
& \quad + \Theta\Bigl(\frac12 - z_l\Bigr)\Theta\Bigl(z_l-\frac13\Bigr) \bigl(C_F^2 B_{q,1} + C_F C_A B_{q,2} + C_F n_f T_F B_{q,3}\bigr) 
\biggr\} \, ,
\end{align}
where
\begin{align}
A_{q,1} &=  4 \left[\frac{\ln(1-z_l)}{1-z_l}\right]_+ + \frac{3}{[1-z_l]_+} +\Bigl(\frac{9}{8}-\frac{\pi ^2}{3}\Bigr) \delta(1-z_l) +\frac{\left(1-2 z_l^2-3 z_l\right)}{2(1-z_l)}\, \ln z_l \nonumber \\
& \quad -\Bigl(z_l-\frac{2}{z_l}+4\Bigr) \ln (1-z_l)-\frac{3}{2}\left(1+\frac{z_l}{2}\right) \, ,\nn \\
A_{q,2} &= -\frac{2 \left(z_l^2+z_l+1\right)}{z_l}\, \ln z_l +\Bigl(z_l+\frac{2}{z_l}-2\Bigr) \ln (1-z_l)+\frac{8 z_l^3+17 z_l^2+26 z_l-40}{12z_l} \, ,\nn \\
A_{q,3} &= \frac{2+5z_l-4z_l^2-4 z_l^3}{3 z_l}+2 (1+z_l) \ln (z_l) \, , \nn \\
B_{q,1} &= -\frac{2\left(3 z_l^2-3 z_l+2\right)}{z_l(1-z_l)}\,\ln (1-2 z_l)+\frac{\left(15 z_l^2-15 z_l+8\right)}{2 z_l(1-z_l)}\,\ln z_l+\frac{3}{2} \ln \Bigl(\frac{1-z_l}{2}\Bigr) \nonumber \\
& \quad +\frac{3 (1-3z_l) (2-z_l)^2}{4z_l(1-z_l)}  \, , \nn \\
B_{q,2} &= \frac{3 z_l-3 z_l^2-2}{z_l(1-z_l)}\,\ln (1-2 z_l)+\frac{-z_l^3+z_l+2}{z_l(1-z_l)}\, \ln z_l - (z_l+4) \ln \Bigl(\frac{1-z_l}{2}\Bigr) \nonumber \\
& \quad +\frac{-45 z_l^6+96 z_l^5+30 z_l^4-352 z_l^3+459 z_l^2-236 z_l+40}{12 (1-z_l)^4 z_l}\, , \nn \\
B_{q,3} &= 2 (1+z_l) \ln\Bigl(\frac{1-z_l}{2z_l}\Bigr)+\frac{18 z_l^6-87 z_l^5+186 z_l^4-152 z_l^3+36 z_l^2+11 z_l-4}{6 z_l(1-z_l)^4} \, .
\label{eq:lJet_q_NNLO_color}
\end{align}
The analogous result for the gluon jet function can be found in appendix~\ref{app:lGluon_nnlo}. At this order there can be as many as three jets in the final state, so the emergence of the new region ${1/3<z_l<1/2}$ is expected.

\begin{figure}[t!]
\centering
\includegraphics[width=0.45\textwidth]{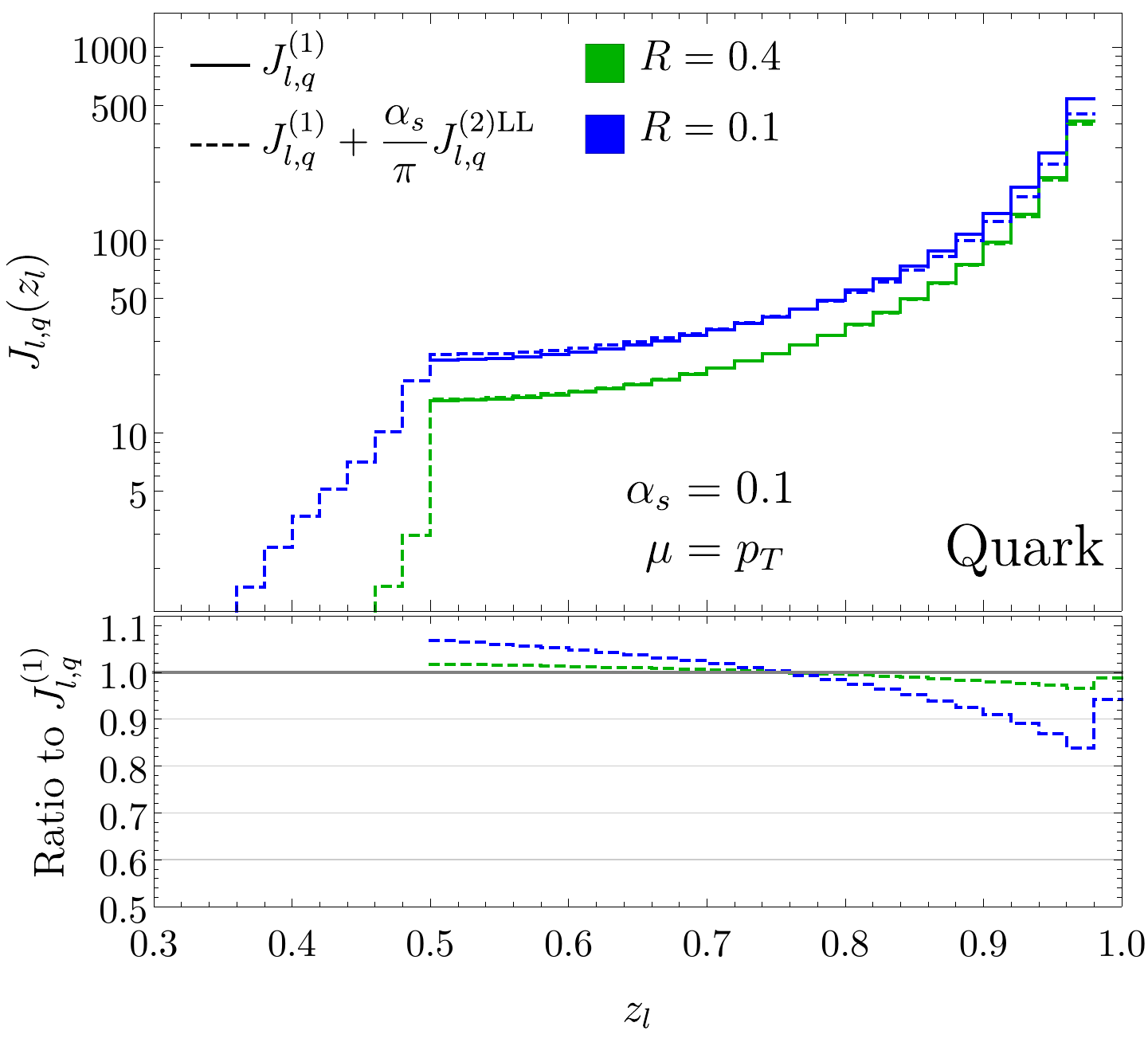} \quad \quad
\includegraphics[width=0.45\textwidth]{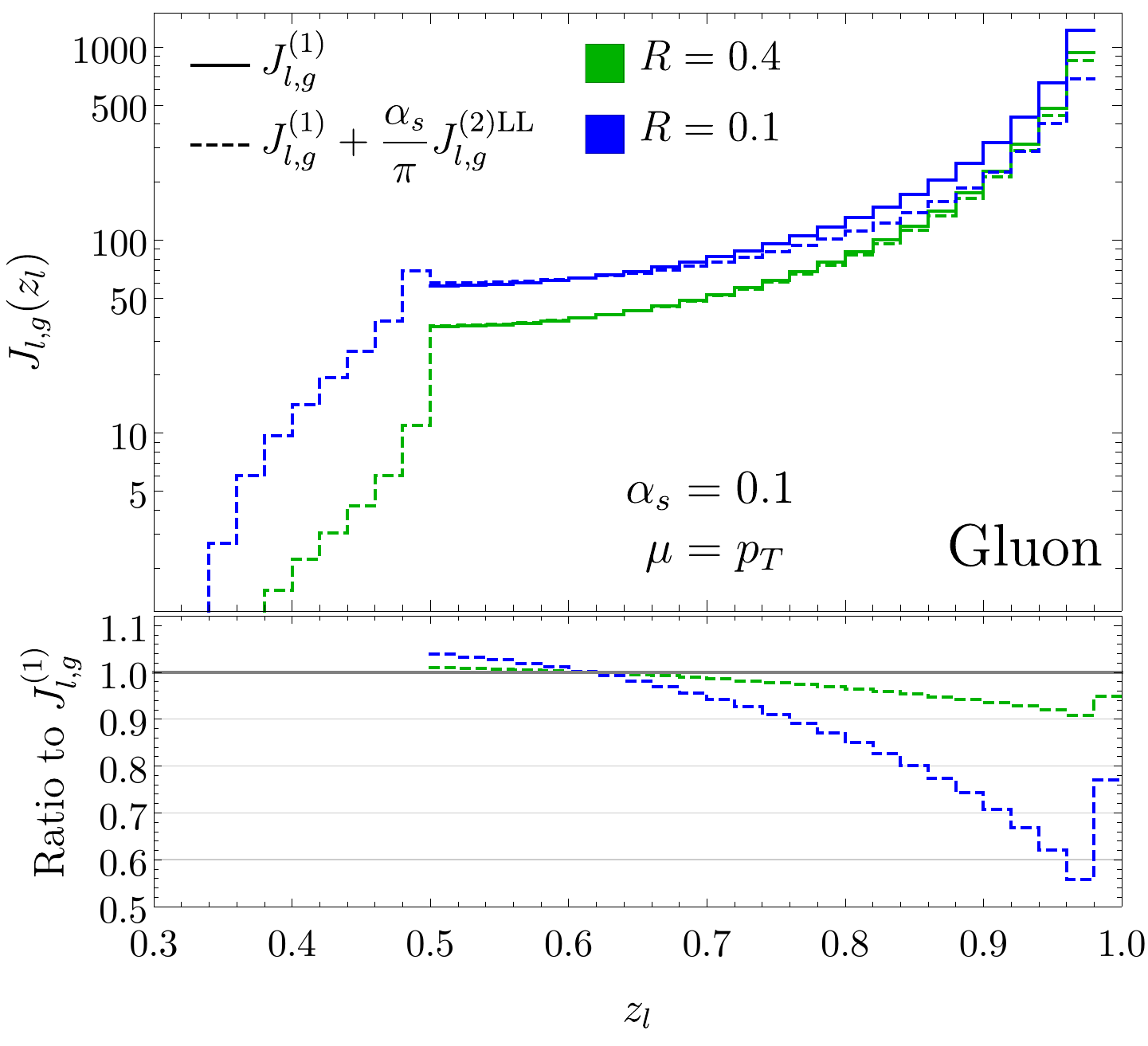}
\caption{Comparison between the NLO without (solid lines) and with the NNLO LL correction (dashed lines) to the quark (left) and gluon (right) leading-jet functions for $R=0.4$ (green lines) and $R=0.1$ (blue lines). The LL terms at NNLO are added to the NLO with the appropriate prefactor $(\alpha_s/\pi)$, see \eq{J_exp}. 
\label{fig:jFn_NNLO_LL}}
\end{figure}
In figure~\ref{fig:jFn_NNLO_LL} we show the effect of the NNLO corrections $J_{l,i}^{(2)\rm{LL}}$ as calculated in \eqs{lJet_q_NNLO}{lJet_g_NNLO} on the NLO quark (left) and gluon (right) leading-jet functions as a function of $z_l$ with $\mu=p_T$. The solid curves in each plot show the full NLO jet functions, while the dashed curves show the effect of adding the NNLO LL terms, weighted appropriately by $\alpha_s/\pi$ for comparison and where we have set $\alpha_s=0.1$. We show results for $R=0.4$ (green) and $R=0.1$ (blue) and the lower panel in each plot displays the ratio compared to the pure NLO prediction for $z_l>1/2$. The LL contributions at NNLO have an integrable divergence as $z_l$ approaches $0.5$ from below. This arises from a soft divergence: the momentum fraction of the softest of the three jets is bound from below by $1-2z_l$, leading to a $\ln(1-2z_l)$ in the $C_F^2$ and $C_F C_A$ color structures in \eq{lJet_q_NNLO_color}. It is clear from the ratio plots, that the NNLO LL corrections for $R=0.4$ have a smaller impact on the quark jet function, where they are below $4\%$, than for the gluon where they approach $10\%$ for larger $z_l$. It is only for more extreme values of $R=0.1$ that the corrections become significant, again particularly for the gluon leading-jet function at large $z_l$. However, given that the LL terms at NLO are not a good approximation for the full NLO jet functions, we might expect the same to occur at NNLO.

Moving on to the subleading-jet function, we find that the $\ord{\al_s^2}$ contribution to the LL evolution is given by
\begin{align}
\label{eq:sJet_NNLO}
J_{s,i}^{(2){\rm LL}}(z_l,z_s, \mu) &= \frac{\beta_0}{4} \ln \Bigl(\frac{\mu}{p_T R}\Bigr) J_{s,i}^{(1){\rm LL}}(z_l,z_s, \mu)  \nn \\
& \quad +\frac{1}{2}\ln^2 \Bigl(\frac{\mu}{p_T R}\Bigr)\Theta\big(z_l-z_s\big) \Theta\big(z_l+2z_s-1\big) \Theta\big(1-z_l-z_s\big) \nn \\
& \quad \, \times \Bigl[F_{i}(z_l,z_s) + F_{i}(z_s,z_l)+ F_{i}(1-z_l-z_s,z_l) \Bigr] \, ,
\end{align}
where
\begin{align} \label{eq:F_i}
F_q(a,b) &= \frac{1}{1-a}\Bigl\{P_{qq}(a)\Bigl[P_{gg}\Bigl(\frac{b}{1-a}\Bigr) + 2n_f P_{qg}\Bigl(\frac{b}{1-a}\Bigr)\Bigr] \nonumber \\
& \quad + P_{qq}(1-a)\Bigl[P_{qq}\Bigl(\frac{b}{1-a}\Bigr) + P_{gq}\Bigl(\frac{b}{1-a}\Bigr)\Bigr] \Bigr\}\, , \nonumber \\
F_g(a,b) &= \frac{1}{1-a}\Bigl\{P_{gg}(a)\Bigl[P_{gg}\Bigl(\frac{b}{1-a}\Bigr) + 2n_f P_{qg}\Bigl(\frac{b}{1-a}\Bigr)\Bigr] \nonumber \\
& \qquad \qquad + 2n_f P_{qg}(a)\Bigl[P_{qq}\Bigl(\frac{b}{1-a}\Bigr) + P_{gq}\Bigl(\frac{b}{1-a}\Bigr)\Bigr] \Bigr\}\, ,
\end{align}
and 
\begin{align}
J_{s,i}^{(1){\rm LL}}(z_l,z_s) = J_{l,i}^{(1){\rm LL}}(z_l)\delta(z_s-(1-z_l)) \, .
\end{align}
Eq.~\eqref{eq:sJet_NNLO} can be interpreted as a perturbative splitting into partons with momentum fractions $a$ and $1-a$, where the parton with $1-a$ subsequently splits into two partons with momentum fractions $b(1-a)$ and $(1-b)(1-a)$. The splitting probabilities are encoded in \eq{F_i}, and the three terms on the last line of \eq{sJet_NNLO} correspond to the different assignments of leading and subleading jets over these three partons. 
We have checked that the integral of \eq{sJet_NNLO} over $z_s$ reproduces $J_{l,i}^{(2){\rm LL}}$ in \eqs{lJet_q_NNLO}{lJet_g_NNLO}.

Up to this point we have only discussed perturbative solutions to the RG equations in \eqs{RG_lJetFn}{RG_slJetFn}, obtaining analytic results up to order $\al_s^2$. The non-linear nature of these equations makes them difficult to solve, and, as has already been suggested, the inclusion of the LL terms alone will not yield a reliable approximation (which will be demonstrated at the level of the cross section in \sec{pheno}). Nonetheless, it is instructive to investigate all-orders solutions to the RGEs derived here.
To this end we introduce the cumulant $\widetilde{J}_{l,i}$ of the leading-jet functions as 
\begin{equation}
\widetilde{J}_{l,i}(a,\mu) = \int_a^1 \! \df z_l \, J_{l,i}(z_l,\mu) \, ,
\end{equation}
from which the original distribution can be easily obtained using
\begin{equation}
J_{l,i}(z_l,\mu) = -\frac{\df}{\df a}\widetilde{J}_{l,i}(a,\mu) \bigg|_{a=z_l} \, .
\end{equation}
The advantage of working with cumulants is the regularization of the distribution-valued contributions to the jet function. 
We can rewrite \eq{RG_lJetFn} in terms of cumulants, yielding
\begin{align} \label{eq:RG_lJet_cumulant}
\mu \frac{\df}{\df \mu} \widetilde{J}_{l,q}(a,\mu) &= \frac{\alpha_s}{\pi} \biggl\{\int_a^1 \! \df z \, \Bigl[ P_{qq}(z) \widetilde{J}_{l,q}\Big(\frac{a}{z},\mu\Big) + P_{gq}(z) \widetilde{J}_{l,g}\Big(\frac{a}{z},\mu\Big)\Bigr] \nn \\
& \quad -\Theta\Big(\frac{1}{2}-a\Big)\int_a^{1-a} \! \df z \, P_{qq}(z) \widetilde{J}_{l,q}\Big(\frac{a}{z},\mu\Big)\widetilde{J}_{l,g}\Big(\frac{a}{1-z},\mu\Big) \biggr\}\, , \nn \\ 
\mu \frac{\df}{\df \mu} \widetilde{J}_{l,g}(a,\mu) &= \frac{\alpha_s}{\pi} \biggl\{\int_a^1 \! \df z \, \Bigl[ \frac12 P_{gg}(z) \widetilde{J}_{l,g}\Big(\frac{a}{z},\mu\Big) + n_f P_{qg}(z) \widetilde{J}_{l,q}\Big(\frac{a}{z},\mu\Big)\Bigr] \nn \\
& \quad -\Theta\Big(\frac{1}{2}-a\Big)\int_a^{1-a} \! \df z \, \Bigl[ \frac12 P_{gg}(z) \widetilde{J}_{l,g}\Big(\frac{a}{z},\mu\Big)\widetilde{J}_{l,g}\Big(\frac{a}{1-z},\mu\Big) \nn \\
& \quad + n_f P_{qg}(z) \widetilde{J}_{l,q}\Big(\frac{a}{z},\mu\Big)\widetilde{J}_{l,q}\Big(\frac{a}{1-z},\mu\Big)\Bigr]\biggr\} \,.
\end{align}
In obtaining this expression, we have made use of the sum rule \eq{sumrule}. It is possible to solve \eq{RG_lJet_cumulant} using a Runge-Kutta algorithm or similar. We note that for $a>1/2$, the above RGEs linearize and become simple convolutions, which can be solved to all orders by making use of a Mellin transform.

\section{Application to Higgs + 1 jet production at the LHC}
\label{sec:pheno}

In this section we present some applications of the (sub)leading-jet function to Higgs + 1 jet production. 
In \sec{NLO_sing} we describe how we obtain the singular NLO cross section (in particular the one-loop hard function) for the jet and Higgs transverse momentum spectrum.
In \sec{NLO_comp} we assess the size of the power corrections to the factorization formula in \eq{fact_sub_nlo} at NLO, and the effect of the $\ln R$ terms at order $\al_s$ and beyond.

\subsection{Construction of NLO predictions}
\label{sec:NLO_sing}

As discussed in \sec{intro}, the subleading-jet function can be used to describe exclusive jet production with a loose veto on the momenta of additional jets. The relevant factorization is given in \eq{fact_sub_nlo}, which we write here as
\begin{align} \label{eq:fact_sub_nlo_ptv}
\frac{\df \sigma^{\rm NLO}_{pp\rightarrow HJ}}{\df p_{T,J}}(p_T^{\rm veto}) &= \frac{\df \sigma^{\rm sing, NLO}_{pp\rightarrow HJ}}{\df p_{T,J}}(p_T^{\rm veto}) +\ord{R^2} \nn \\
&= \sum_{i,j} \frac{\df \tilde{\sigma}_{pp \rightarrow Hij}}{\df p_{T,i}\df p_{T,j}} \otimes J_{s,i} \otimes J_{s,j} +\ord{R^2} \,.
\end{align}
We denote the factorized (singular) expression for the cross section by $\df \sigma^{\rm{sing}}$ and on the last line  introduce a shorthand notation that will be convenient for our discussion below. 
Expanding the right-hand side of \eq{fact_sub_nlo_ptv} to NLO, we obtain
\begin{align}
\label{eq:NLO_sing_ptJ}
\frac{\df \sigma^{\rm sing, NLO}_{pp\rightarrow HJ}}{\df p_{T,J}}(p_T^{\rm veto}) &= 
\sum_{i} \frac{\df \tilde{\sigma}^{(0)}_{pp \rightarrow Hi}}{\df p_{T,i}} \!\otimes\!  J_{s,i}^{(0)} \nonumber \\ 
& \quad + \frac{\alpha_s}{\pi} \biggl[\sum_{i} \frac{\df \tilde{\sigma}^{(0)}_{pp \rightarrow Hi}}{\df p_{T,i}} \!\otimes\! J^{(1)}_{s,i} 
\!+\! \sum_{i,j} \frac{\df \tilde{\sigma}^{(1)}_{pp \rightarrow Hij}}{\df p_{T,i}\df p_{T,j}}  \!\otimes\! J_{s,i}^{(0)} \!\otimes\!  J_{s,j}^{(0)} \biggr].
\end{align}
As noted below \eq{fact_lo}, $\df \tilde{\sigma}^{(0)}_{pp\rightarrow Hi}$ is simply the LO partonic cross section for the production of a Higgs boson and a parton $i$. Since the LO subleading-jet functions are simply delta functions, convolutions involving them are trivial and the first term on the right-hand side is just the LO cross section for H+jet production $\df \sigma^{\rm{LO}}_{pp\rightarrow HJ}$.
While $\df \tilde{\sigma}^{(1)}_{pp \rightarrow Hij}$ can be computed directly from the hard scattering, here we instead extract it from the known full NLO cross section. It can be extracted from \eq{NLO_sing_ptJ}  by replacing $\df \sigma^{\rm sing,NLO}$ with $\df \sigma^{\rm NLO}$ at small values of $R$ (we took $R=0.05$) to minimize the $\ord{R^2}$ corrections. Since $\df \tilde{\sigma}^{(1)}_{pp \rightarrow Hij}$ is independent of $R$, it can be used for other values of $R$ once it has been extracted. Explicitly
\begin{align}
\label{eq:NLO_extract}
\frac{\alpha_s}{\pi} \frac{\df \tilde{\sigma}^{(1)}_{pp \rightarrow Hij}}{\df p_{T,i}\df p_{T,j}}  \!\otimes\! J_{s,i}^{(0)} \!\otimes\!  J_{s,j}^{(0)} &= \frac{\df \sigma^{\rm NLO}_{pp\rightarrow HJ}}{\df p_{T,J}}(p_T^{\rm veto})\bigg|_{R=0.05} \nn \\ & \quad - \sum_{i}\biggl[ \frac{\df \tilde{\sigma}^{(0)}_{pp \rightarrow Hi}}{\df p_{T,i}} \!\otimes\!  J_{s,i}^{(0)}+\frac{\alpha_s}{\pi} \frac{\df \tilde{\sigma}^{(0)}_{pp \rightarrow Hi}}{\df p_{T,i}} \!\otimes\! J^{(1)}_{s,i}\biggr|_{R=0.05} \biggr].
\end{align}
In \eq{NLO_extract} the left hand side is essentially $\df \tilde{\sigma}^{(1)}$ since the leading order jet functions with which it is convolved are delta functions.

We can also use the (sub)leading-jet functions to investigate the transverse momentum spectra of the colorless object (in our case the Higgs boson) recoiling against the jets. To derive the corresponding expression, we first make all cross sections in \eq{NLO_sing_ptJ} also differential in the Higgs transverse momentum $p_{T,H}$ and then integrate over $p_{T,J}$. Specifically,
\begin{align}
\label{eq:NLO_sing_pth}
\frac{\df \sigma^{\rm{sing,NLO}}_{pp\rightarrow HJ}}{\df p_{T,H}} &= \frac{\df \sigma^{\rm{LO}}_{pp\rightarrow HJ}}{\df p_{T,H}} + 
 \frac{\alpha_s}{\pi} \biggl[
\sum_{i,j} \int \! \df p_{T,J} \, \frac{\df \tilde{\sigma}^{(1)}_{pp \rightarrow Hij}}{\df p_{T,i}\df p_{T,j} \df p_{T,H}} \otimes J^{(0)}_{s,i} \otimes J^{(0)}_{s,j} \nn \\
& \quad + \sum_{i} \int \! \df p_{T,J} \, \frac{\df \tilde{\sigma}^{(0)}_{pp \rightarrow Hi}}{\df p_{T,i} \df p_{T,H}} \otimes J^{(1)}_{s,i}\biggr] \, .
\end{align}
At LO, $p_{T,H}=p_{T,J}$, so the first term on the right-hand side is the same as for the singular cross section with the $p_{T,J}$ measurement. In the last term of \eq{NLO_sing_pth}, $p_{T,H} = p_{T,i}$ allowing us to rewrite this term as 
\begin{align}
\label{eq:H0J1_pth}
& \sum_i \int \! \df p_{T,J} \, \frac{\df \tilde{\sigma}^{(0)}_{pp \rightarrow Hi}}{\df p_{T,i} \df p_{T,H}} \int \! \df z_{l,i} \df z_{s,i} J_{s,i}^{(1)}(z_{l,i},z_{s,i}) \delta(p_{T,J}-z_{l,i} p_{T,i})
\Theta(p_{T}^{\rm{veto}}-z_{s,i} p_{T,i}) \nn \\
& \quad = \sum_i \frac{\df \tilde{\sigma}^{(0)}_{pp \rightarrow Hi}}{\df p_{T,H}}  \int_0^1 \! \df z_{l,i}\df z_{s,i} \, J_{s,i}^{(1)}(z_{l,i},z_{s,i}) \Theta(p_{T}^{\rm{veto}}-z_s p_{T,H})\, ,
\end{align}
The final ingredient required for the Higgs $p_T$ spectrum is the second term on the right-hand side of \eq{NLO_sing_pth}. Since this term is independent of $R$, it can again be extracted from the full NLO result at small jet radius $R$.

\subsection{Comparison with full NLO predictions}
\label{sec:NLO_comp}

We will now perform a phenomenological study using the cross sections derived in \sec{NLO_sing}.
In our numerical results, we take the Higgs mass $m_H = 125$~GeV and vacuum expectation value $v=246$~GeV. We set the factorization and renormalization scales equal to the transverse mass of the Higgs boson, $\mu_F=\mu_R=m_{T,H}=\sqrt{m_H^2+p_{T,H}^2}$, and restrict the jets to the pseudo-rapidity range $|\eta_J|<4.5$. Predictions for NLO distributions are obtained from MCFM-8.3~\cite{Campbell:1999ah,Campbell:2011bn,Campbell:2015qma}, which makes use of virtual matrix elements calculated in refs.~\cite{Schmidt:1997wr,Ravindran:2002dc}, with the anti-$k_T$ jet clustering algorithm~\cite{Cacciari:2008gp}. We employ the NNLO PDF4LHC15 parton distributions~\cite{Butterworth:2015oua}, within the LHAPDF-6.2.3 framework~\cite{Buckley:2014ana}. We have chosen to use NNLO PDFs to match the order of our results at the end of this section which include $\alpha_s^2 \ln^2 R$ corrections. This avoids effects from using PDFs at different orders, which would otherwise obscure the size of the corrections coming from the $\ln R$ terms, which we are ultimately interested in.

As a first step, it is insightful to check how good of an approximation the factorized differential cross sections $\df \si^{\rm sing,NLO}$ in \eqs{NLO_sing_ptJ}{NLO_sing_pth} are to the full NLO predictions.
\begin{figure}[t!]
\centering
\begin{align*}
\includegraphics[width=0.495\textwidth]{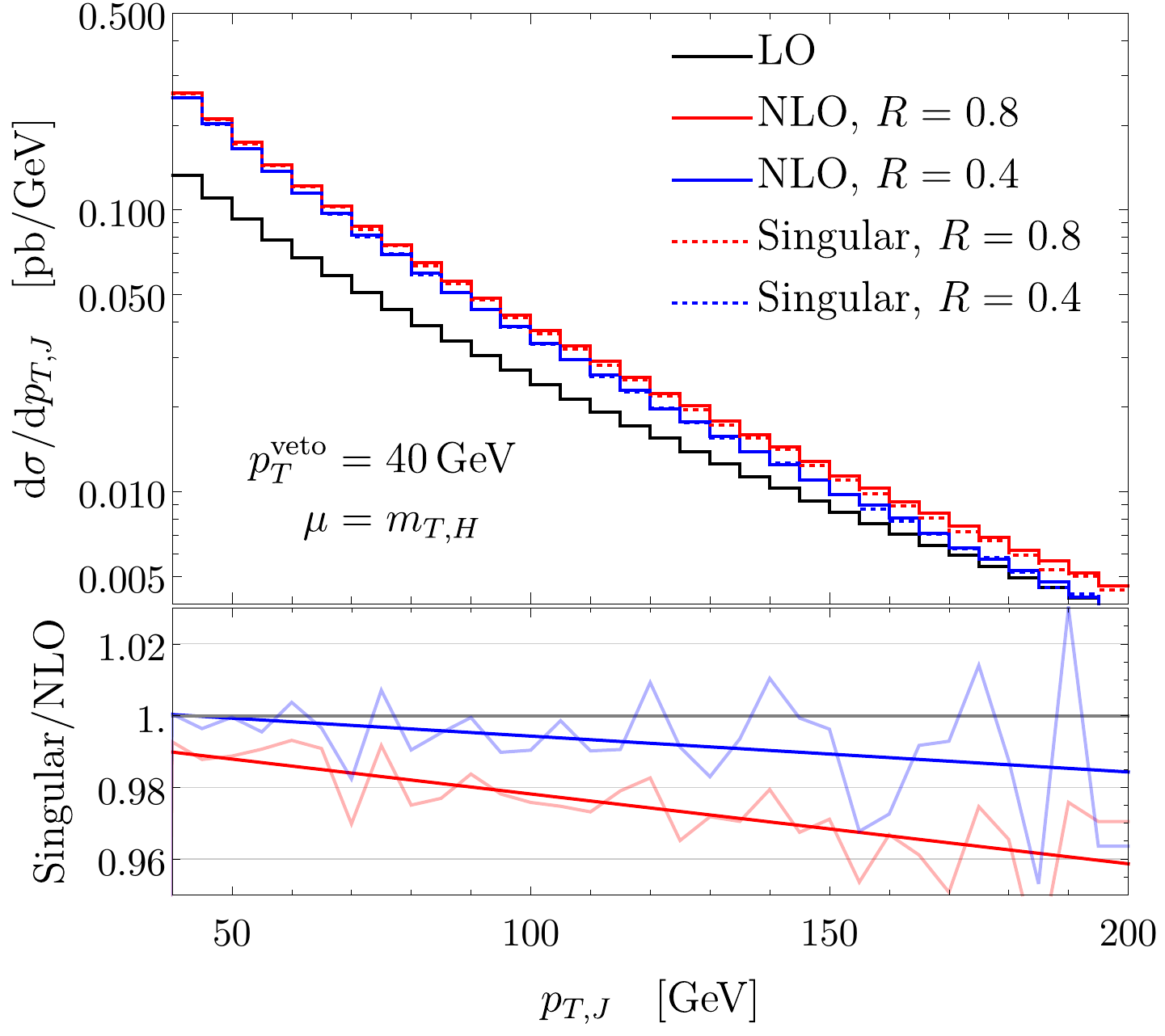}
\includegraphics[width=0.495\textwidth]{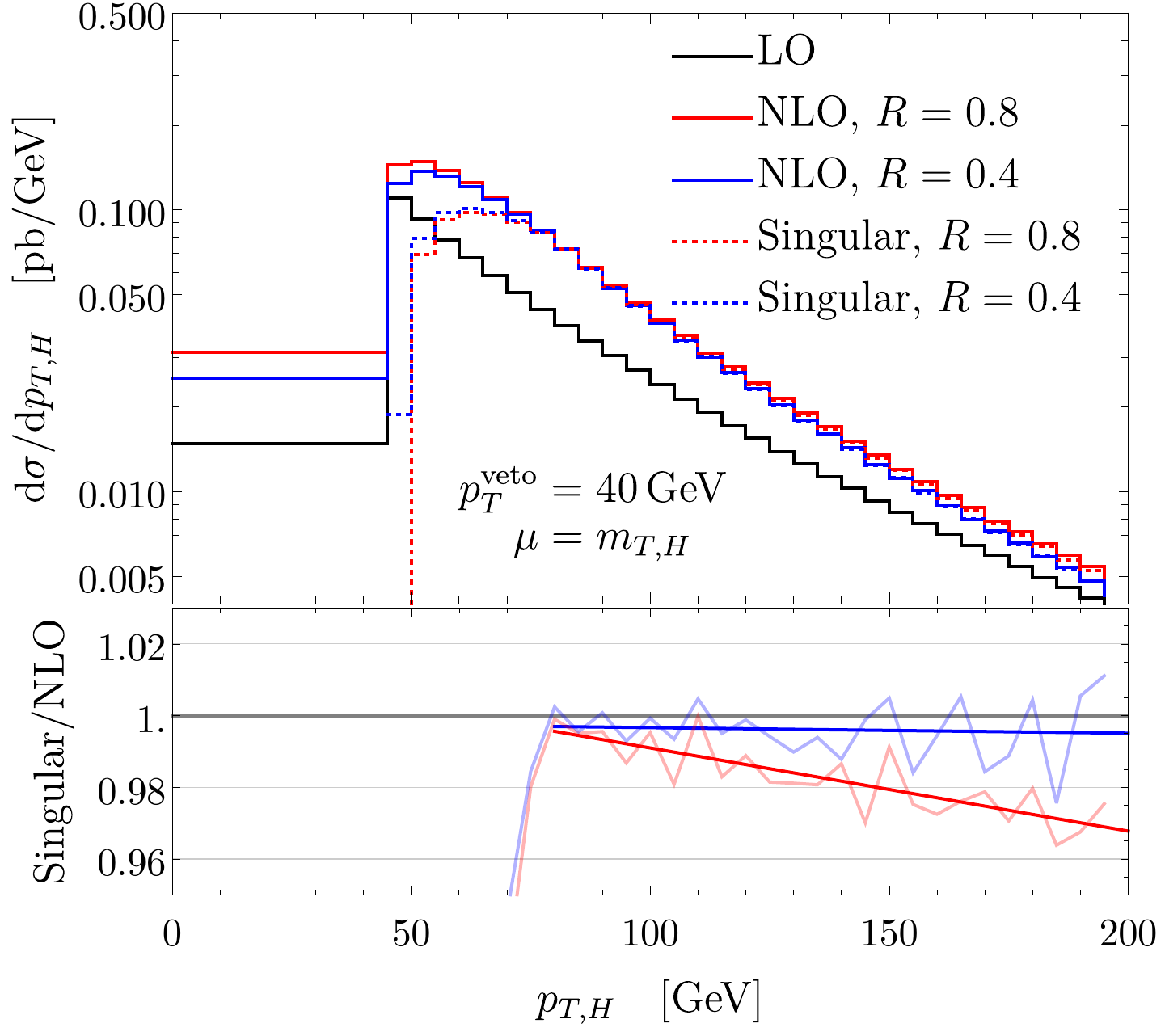}
\end{align*}
\caption{\label{fig:nloVsing}
The full NLO prediction (solid lines) for exclusive Higgs + 1 jet production for the jet (left) and Higgs (right) transverse momentum distributions with $p_T^{\rm{veto}}=40$ GeV compared to the NLO singular prediction in \eq{NLO_sing_ptJ} (dotted lines). The anti-$k_T$ jet algorithm with $R=0.8$ (red) and $R=0.4$ (blue) is used. The LO cross section (solid gray) is shown for comparison. The lower panels show the ratio of the singular and NLO predictions (more transparent line) as well as a linear fit of this ratio to clarify the trend (more opaque).}
\end{figure}
To this end we show in \fig{nloVsing} the differential cross section for the jet (left) and Higgs (right) transverse momentum distribution in exclusive Higgs + 1 jet production. In the upper panels the full NLO predictions for $R=0.8$ and $R=0.4$ (solid red and blue lines) and the singular cross section (dashed lines) are shown. For comparison the leading order cross section (solid black line) is shown. Taking $p_T^{\rm{veto}}=40$ GeV ensures that there is not too large a hierarchy between $p_T^{\rm{veto}}$ and other hard scales, i.e.~$m_H \sim p_{T,J} \sim p_{T,H} \sim p_T^{\rm{veto}}$, since the calculation would otherwise require additional resummation. While $p_{T,J}\geq p_{T}^{\rm veto}$, the Higgs distribution extends down to $p_{T,H}=0$ GeV. However, the veto at $40$ GeV gives rise to a ``Sudakov shoulder''~\cite{Catani:1997xc} in the $p_{T,H}$ spectrum around $p_T^{\rm veto}$ and so we therefore use one large bin accounting for all of the cross section at $p_{T,H} < 45$ GeV.\footnote{Any detailed predictions in this low-energy part of the spectrum would anyway be outside the range of validity of our factorization. A double differential resummation in this region is performed in~\cite{Monni:2019yyr}.} It is clear from the figure that the expansion in small $R$ works well, as evidenced by the ratio plots in the lower panels of the figure. In the lower panels the lighter lines show the ratio $\df \sigma^{\rm{sing,NLO}}/ \df \sigma^{\rm{NLO}}$, with fluctuations due to the limited statistics of the Monte Carlo integration, while the thicker lines display a linear fit in order to highlight the trend. For the Higgs distribution we perform this fit only for $p_{T,H}>80$ GeV to avoid the influence of the Sudakov shoulder. While both values of the jet radius considered are in excellent agreement with the full NLO prediction, delivering predictions typically within a few percent of the exact NLO, factorized predictions for smaller radii give a more accurate prediction of the full NLO, as expected. The validity of the small $R$ approximation for relatively large values of $R$ has been noted before in e.g.~refs.~\cite{Mukherjee:2012uz,Dasgupta:2016bnd}.
\begin{figure}[t!]
\centering
\includegraphics[width=0.49\textwidth]{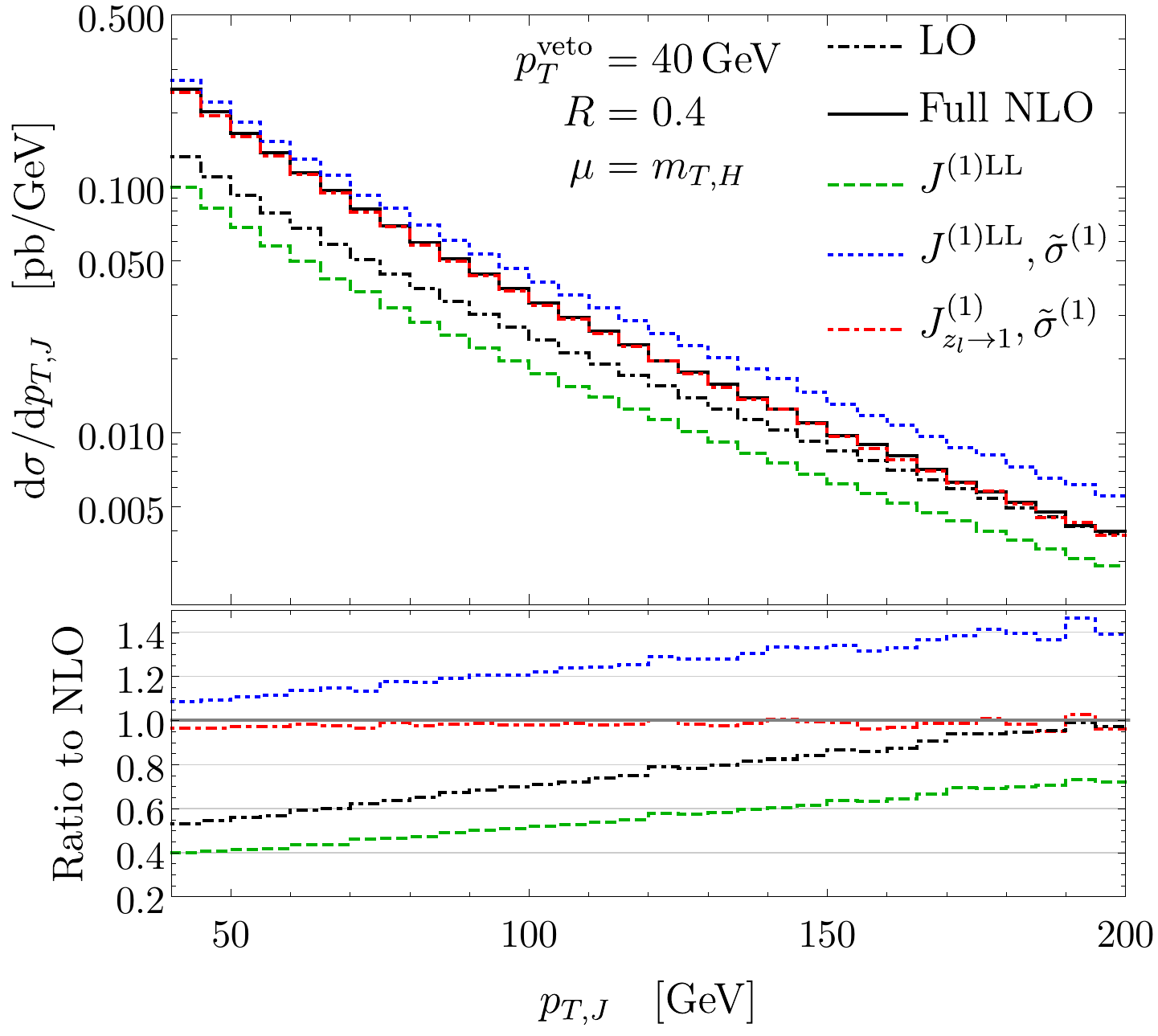}
\includegraphics[width=0.49\textwidth]{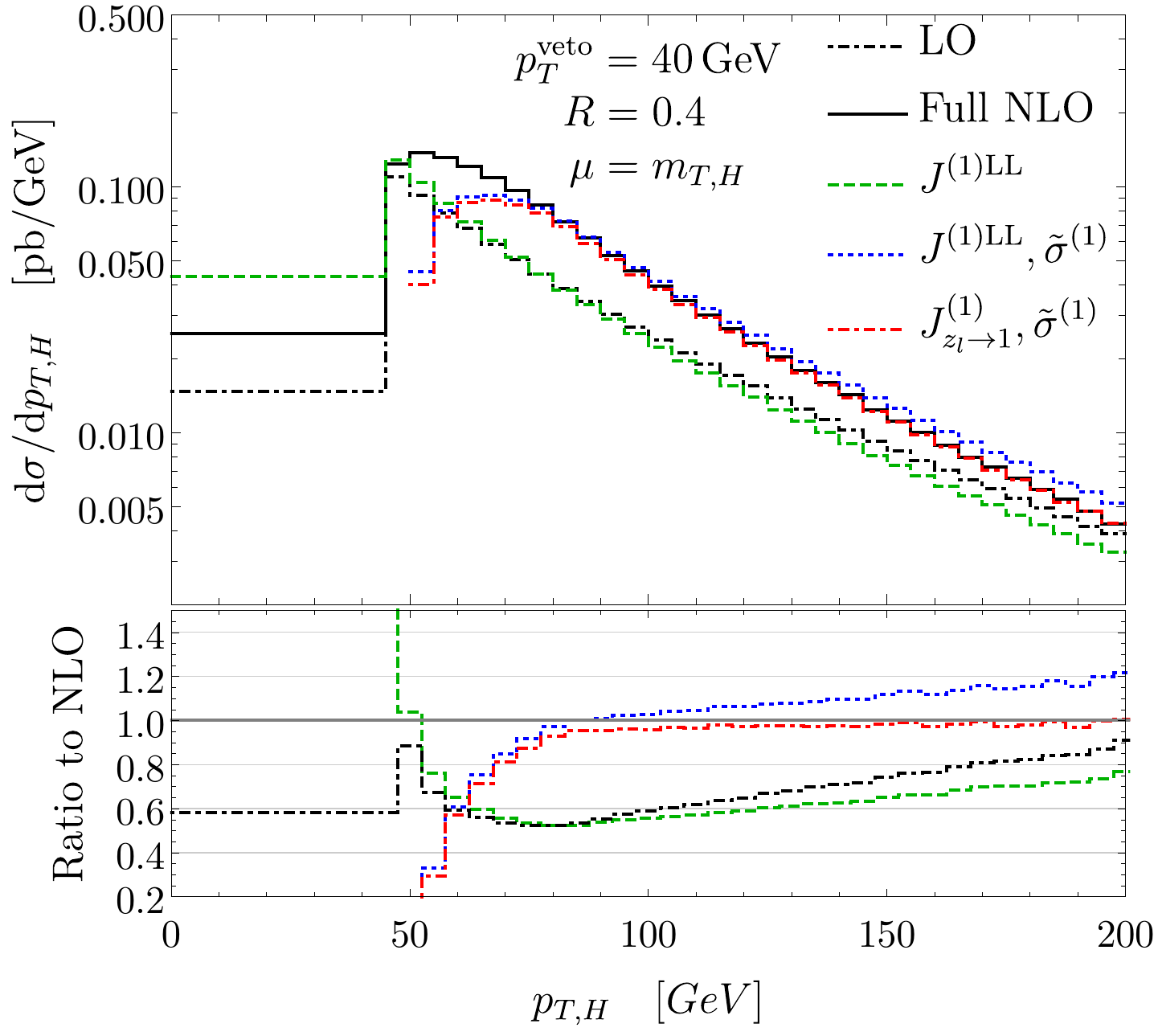}
\caption{\label{fig:nloVsing_part}
The $p_{T,J}$ (left) and $p_{T,H}$ (right) distributions in exclusive Higgs + 1 jet production at LO (dot-dashed black) and NLO (solid black) with $R=0.4$ and $\mu=m_{T,H}$. Also shown are predictions from the singular cross section using the LL (blue dashed) and soft (red dot-dashed) approximations of the jet functions, as well as the LO cross section augmented only with the $\al_s \ln R$ terms (green dashed).}
\end{figure}

Given that the collinear approximation works so well, we next investigate how much of this is is captured by the $\al_s \ln R$ terms in the singular cross section. In \fig{nloVsing_part} we show the $p_{T,J}$ (left) and $p_{T,H}$ (right) distributions for $R=0.4$, keeping various components of the singular cross section. Shown are predictions obtained by simply adding the LL terms at $\ord{\al_s}$ to the LO cross section (green dashed) as well as the NLO singular cross section, \eq{NLO_sing_ptJ}, using the LL (blue dotted) and soft (red dot-dashed) approximation of the full NLO jet function. The LO (dot-dashed black line) and full NLO (solid black line) contribution are shown for comparison. The bottom panel of each plot displays the ratio of each prediction to the full NLO result.
For both the $p_{T,J}$ and $p_{T,H}$ distributions, we see that simply augmenting the LO cross section with the $\al_s \ln R$ terms (green dashed line) gives results which severely underestimate the exact NLO cross section. Indeed, from the ratio plots we can see that it only captures between $40$ -- $70\%$ of the cross section for the $p_{T,J}$ distribution depending on the $p_{T,J}$ bin and does only slightly better (away from the Sudakov shoulder) for $p_{T,H}$. The use of the LL approximation of the NLO jet function in singular cross section prediction (blue dashed lines) leads to a much better approximation of the full NLO prediction, though the results are typically overestimated and only get worse for larger transverse momentum. This is line with what we observed in \fig{jFn_NLO}, where the $\al_s \ln R$ term accounts for only $30$ -- $40\%$ of the full NLO jet function at $R=0.4$ across the full range of $z_l$.
In \fig{jFn_NLO} we also noted that the soft limit $z_l \rightarrow 1$ of the jet function in \eq{J_z1} gives a much more faithful approximation of the full result. 
Using this soft approximation of the jet function in the NLO singular cross sections \eqs{NLO_sing_ptJ}{NLO_sing_pth} (red dot-dashed lines) yields results which are almost identical to the complete NLO results for $p_{T,J}$ and, outside the influence of the Sudakov shoulder, for $p_{T,H}$.

The overriding conclusion of these studies is that while the $R \ll 1$ limit works extremely well, even for ``large'' $R$ values (e.g.~$R = 0.8$), the dominant effects here are not given by the $\alpha_s \ln R$ corrections to the LO result.
This is perhaps unsurprising given that the NLO (and higher) QCD corrections to Higgs + jet production are sizable, so that we might expect large corrections from the NLO hard function $\df \tilde{\sigma}^{(1)}$. However, it is also clear that the jet function gives an important contribution, of which the $\al_s \ln R$ term provides a poor approximation. Interestingly though, since the contribution from the $\alpha_s \ln R$ terms is negative (the cross section is smaller than the LO result), using smaller values of $R$ only makes this a worse approximation to the full NLO result. 

\begin{figure}[t!]
\centering
\includegraphics[width=0.495\textwidth]{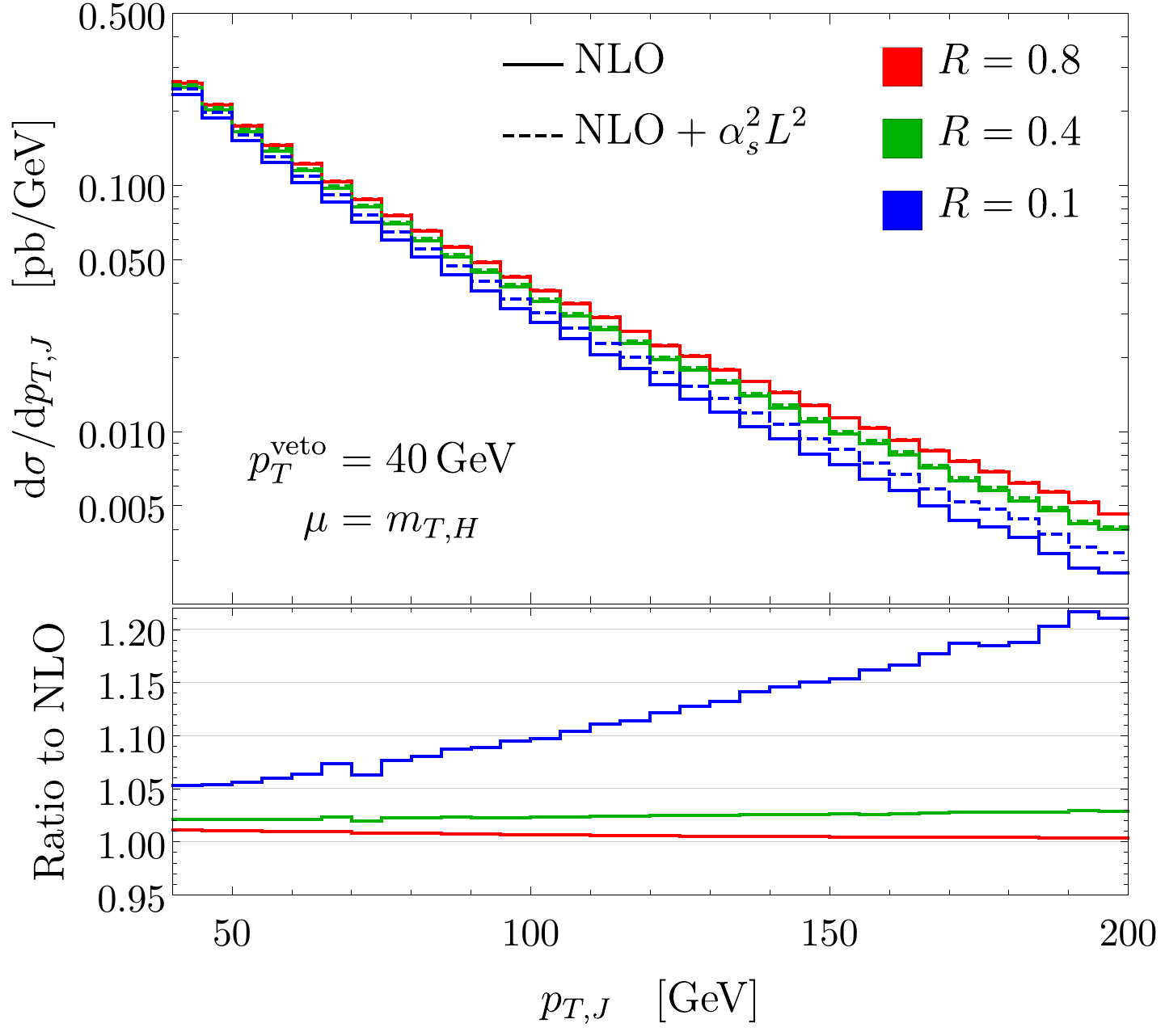}
\includegraphics[width=0.495\textwidth]{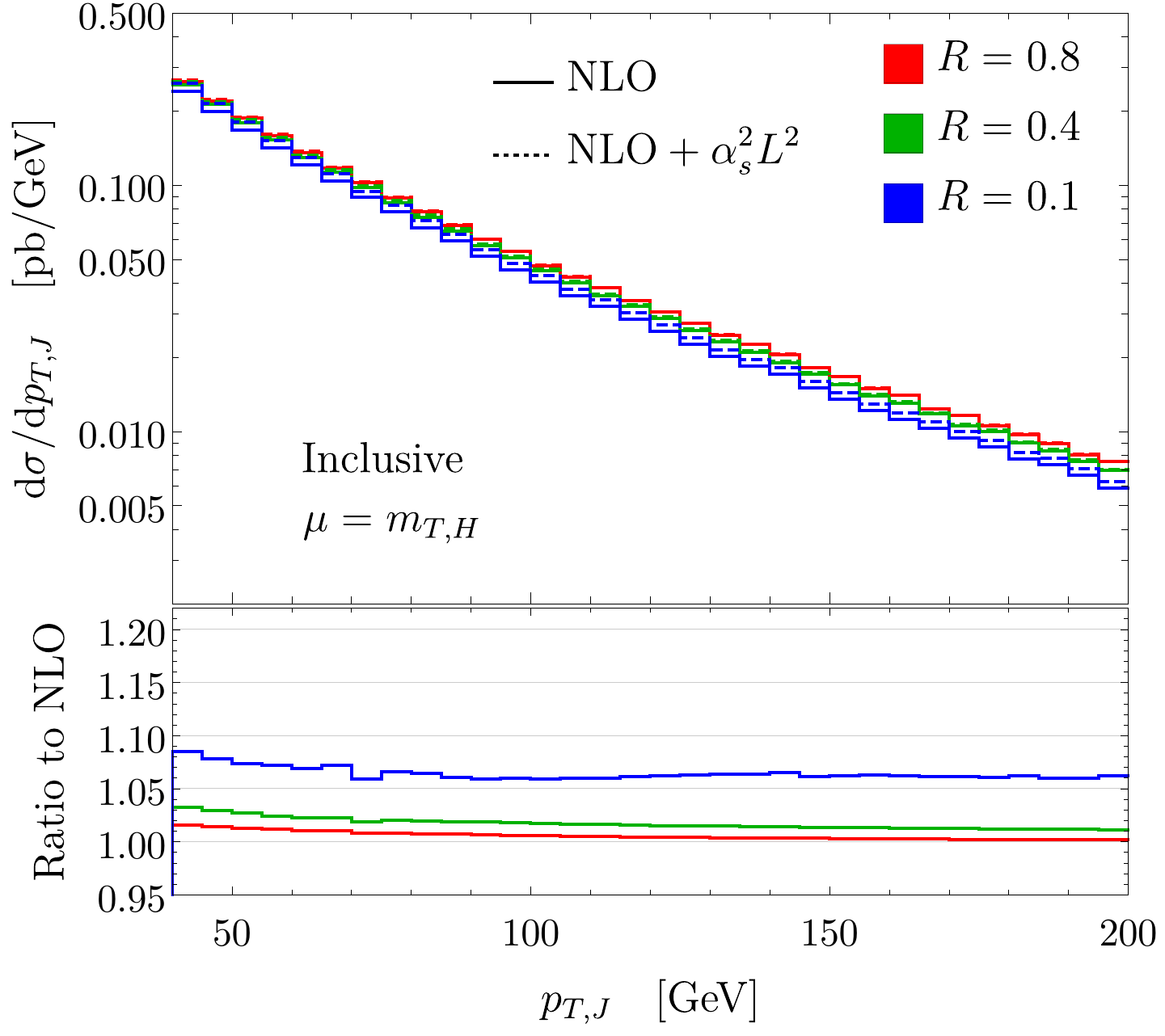}
\caption{\label{fig:nloVaL2_jet}
The Higgs + 1 jet cross section differential in the jet transverse momentum at NLO (solid lines) and NLO + $\alpha_s^2 \ln^2 R$ corrections (dashed lines) for $R=0.8, 0.4, 0.1$ (red, green, blue). The left plot shows the exclusive cross section with a loose $p_T^{\rm veto}=40$ GeV, while the right plot shows the inclusive cross section. The lower panels show the ratio of the NLO + $\alpha_s^2 \ln^2 R$ predictions with the corresponding NLO ones.}
\end{figure}
Finally, even though we have established that simply augmenting the LO results with $\alpha_s \ln R$ corrections does not provide a good approximation of the full NLO, we can still assess how big an impact the $\alpha_s^2 \ln^2 R$ terms have.
In figure~\ref{fig:nloVaL2_jet} we show a comparison between the exact NLO prediction (solid lines) and the NLO prediction plus the leading $\alpha_s^2 \ln^2 R$ terms (dashed lines) for the jet transverse momentum spectrum at varying values of $R$. In the left plot, we show the exclusive cross section with a loose jet veto $p_{T}^{\rm veto}=40$~GeV where we see that for moderate values of $R\sim 0.8, 0.4$ the effect of the $\ln^2 R$ terms is minimal. As can be seen in the lower panel, the effect is around $2$--$3\%$ for $R=0.4$ and at most $1\%$ for $R=0.8$. It is only for the more extreme value of $R=0.1$ that the $\ln^2 R$ terms start to have a large impact, especially at large transverse momentum. The more pronounced $p_{T,J}$ dependence at $R=0.1$ has a simple explanation:
the transverse momentum dependence of the $\ln^2 R$ term is similar to that of the LO cross section, which falls off slower than the exclusive NLO cross section, an effect which is enhanced for smaller $R$ values (c.f. $R=0.8$ and $R=0.4$ predictions in figure~\ref{fig:nloVsing}). As such, the contribution from the $\alpha_s^2 \ln^2 R$ terms, as a fraction of the NLO predictions, grows at larger $p_{T,J}$ for these smaller values of $R$. 
 While this is potentially an interesting indication of where $\ln^2 R$ terms could become large enough to warrant resummation, it is also well known that the NNLO QCD corrections to Higgs plus jet production cross sections lead to large $K$-factors~\cite{Chen:2014gva,Boughezal:2015aha}. We would therefore also expect large corrections from the $\df \tilde{\sigma}^{(2)}$ term in the factorization formula, as well as potentially large contributions from $\df \tilde{\sigma}^{(1)} \otimes J^{(1)}$.
The plot on the right of figure~\ref{fig:nloVaL2_jet} instead displays the inclusive cross section, where we also assess the impact of the $\alpha_s^2 L^2$ terms. In this instance the effect of the $\alpha_s^2 L^2$ terms is relatively unchanged for $R=0.4$, while the enhancement seen for $R=0.1$ in figure~\ref{fig:nloVaL2_jet} is reduced to $\lesssim 6\%$ enhancement for most of the spectrum. Thus, the simple inclusion of these terms cannot lead to the kind of K-factors needed to approximate the full NNLO inclusive cross section.

\section{Conclusion}
\label{sec:conc}

In this paper we have explored the validity of the small $R$ approximation and the impact of $\ln R$ terms, developing the general framework and showing results for Higgs plus jet production. To describe the transverse momentum spectrum of the hardest jet in the final state (possibly with a loose veto on additional jets), we introduced (sub)leading-jet functions that describe the momentum fractions of the (second) hardest jet produced by a hard parton. We have calculated these jet functions for quarks and gluons to next-to-leading order in QCD. This calculation revealed that these jet functions are poorly approximated when only the $\al_s \ln R$ terms are retained, even for $R=0.1$, but are well described by their soft limit.
Using a recursive parton shower picture, we derived the renormalization group equations for the jet functions at leading logarithmic accuracy. These RGEs have an interesting non-linear structure, and so their all-orders solution lends itself to more numerical methods. Nonetheless we used it to produce analytical results for the LL terms at NNLO.

We assessed the impact of these jet functions at the cross section level in the concrete example of exclusive Higgs + 1 jet production with a loose transverse momentum veto on additional jets. We confirmed the validity of the collinear factorization as an approximation to the full cross section at NLO, even for the large value of $R=0.8$. Furthermore, we showed that the NLO cross section is not well approximated by simply augmenting the LO result with the $\alpha_s \ln R$ terms at NLO. The NLO corrections to the hard scattering are substantial, as is well-known for Higgs production, but also the other contributions to the jet function cannot be neglected. Finally, using our analytic results for the $\al_s^2 \ln^2 R$ contribution to the NNLO jet functions, we investigate their effect on the cross section. We find that the impact of these logarithms lies at the few percent level, except for rather small values of $R$ in exclusive cross sections.

In conclusion, we have found that for describing the leading jet, the collinear approximation also works rather well, even for rather large values of $R$. On the other hand, the corresponding $\ln R$ terms in the cross section are not particularly large, suggesting that their resummation is of limited importance. This can of course change as jet substructure techniques utilize subjets with smaller radii, requiring the role of higher order subjet radius logarithms to be reassessed. This may be particularly interesting for track-based observables, where small angular scales are experimentally accessible.

\acknowledgments
This work is supported by the ERC grant ERC-STG-2015-677323, and the D-ITP consortium, a program of the Netherlands Organization for Scientific Research (NWO) that is funded by the Dutch Ministry of Education, Culture and Science (OCW).

\appendix

\section{Higher order solution for gluon leading-jet function}
\label{app:lGluon_nnlo}
In this appendix we present the LL solution for the gluon leading-jet function at order $\al_s^2$
\begin{align}
\label{eq:lJet_g_NNLO}
J_{l,g}^{(2){\rm LL}}(z_l,\mu) &= \ln^2\Bigl(\frac{\mu}{p_T R}\Bigr)\bigg\{\frac{\beta_0}{4}\, \Theta\Bigl(z_l-\frac12\Bigr) \bigl[P_{gg}(z_l)+2n_f P_{qg}(z_l)\bigr] \nn \\
& \quad + \Theta\Bigl(z_l-\frac12\Bigr)\bigl(C_A^2 A_{g,1} + C_A n_f T_F A_{g,2} + C_F n_f T_F A_{g,3} + n_f^2 T_F^2 A_{g,4}\bigr) \nn \\
& \quad + \Theta\Bigl(\frac12-z_l\Bigr)\Bigl(z_l-\frac13\Bigr)\bigl(C_A^2 B_{g,1} + C_A n_f T_F B_{g,2} + C_F n_f T_F B_{g,3} \bigr) 
\bigg\} \, ,
\end{align}
where
\begin{align}
A_{g,1} &= 4 \left[\frac{\ln(1-z_l)}{1-z_l}\right]_+ + \frac{11}{3}\frac{1}{[1-z_l]_+} +\left(\frac{121}{72} - \frac{\pi^2}{3}\right)\delta(1-z_l) + \frac{11 z_l^3-7 z_l^2-4 z_l-11}{3 z_l} \nonumber \\
& \quad -\frac{4 \left(z_l^3-z_l^2+2 z_l-1\right)}{z_l} \ln (1-z_l) -\frac{2
   \left(z_l^4-4 z_l^3+3 z_l^2+1\right)}{(1-z_l) z_l}\ln (z_l)   \, , \nn \\
A_{g,2} &=  -\frac{4}{3}\frac{1}{[1-z_l]_+}-\frac{11}{9}\delta(1-z_l) + \frac{1}{6} \left(-32 z_l^2+18 z_l+33\right) 
\nonumber \\
& \quad 
+2 \left(2 z_l^2-2 z_l+1\right) \ln (1-z_l)+2 (4 z_l+1) \ln z_l\, , \nn \\
A_{g,3} &= 2 \left(2 z_l^2-2 z_l+1\right) \ln (1-z_l)-\left(4 z_l^2-4z_l-1\right) \ln (z_l)-\frac{8 z_l^3-6 z_l^2-3z_l-8}{6 z_l} \, , \nn \\
A_{g,4} &= \frac{2}{9} \delta(1-z_l)-\frac{2}{3} \left(2 z_l^2-2 z_l+1\right) , \nn \\
B_{g,1} &= \frac{2\left(3 z_l^4-8 z_l^3+9 z_l^2-4z_l+3\right)}{z_l(1-z_l)}\ln z_l -\frac{6 \left(z_l^2-z_l+1\right)^2 }{z_l(1-z_l)}\ln (1-2 z_l) 
 -4 (1+z_l) \ln \Bigl(\frac{1-z_l}{2}\Bigr) \nonumber \\
& \quad +\frac{-270 z_l^7+1206 z_l^6-2379 z_l^5+2850 z_l^4-2482 z_l^3+1572z_l^2-593z_l+88}{12z_l(1-z_l)^4} \, , \nn \\
B_{g,2} &=  -4z_l(3-z_l) \ln (z_l) -2 \left(2 z_l^2-2 z_l+1\right) \ln(1-2 z_l) +2 (4z_l+1) \ln \Bigl(\frac{1-z_l}{2}\Bigr) \nonumber \\
& \quad +\frac{90 z_l^7-408 z_l^6+741 z_l^5-652 z_l^4+320 z_l^3-120 z_l^2+45z_l-8}{6z_l(1-z_l)^4} \, , \nn \\
B_{g,3} &= \left(8 z_l^2-8 z_l+1\right) \ln z_l -4 \left(2 z_l^2-2 z_l+1\right) \ln (1-2 z_l) +3 \ln \Bigl(\frac{1-z_l}{2}\Bigr)\nonumber \\
& \quad -\frac{180 z_l^4-258z_l^3+135 z_l^2-47 z_l+8}{6 z_l(1-z_l)}\, .
\end{align}

\bibliographystyle{JHEP}
\bibliography{biblio}

\end{document}